\documentclass[onecollarge,final]{svjour3}                     
\usepackage{graphicx}
\usepackage{appendix}
\usepackage{amsmath, amssymb}
\usepackage{enumerate}
\usepackage{datetime2}
\usepackage{xcolor}

\usepackage{mathtools}
\usepackage{url}

\usepackage{subfig}

\usepackage{tikz}\usetikzlibrary{decorations.markings,patterns,calc}

\newcommand{\be}{\begin{equation}}
\newcommand{\ee}{\end{equation}}
\newcommand{\bea}{\begin{eqnarray}}
\newcommand{\eea}{\end{eqnarray}}

\newcommand{\Tr}{\mathrm{Tr}\,}

\newcommand{\erf}{\mathrm{erf}}
\newcommand{\erfc}{\mathrm{erfc}}

\newcommand{\dd}{{\rm d}}

\newcommand{\ed}{{\rm e}}

\newcommand{\Gen}{{\cal G}}
\newcommand{\lambdaL}{{\lambda,\scriptscriptstyle{L}}}
\newcommand{\lambdaR}{{\lambda,\scriptscriptstyle{R}}}
\newcommand{\mL}{m_{\scriptscriptstyle{L}}}
\newcommand{\mR}{m_{\scriptscriptstyle{R}}}
\newcommand{\mRt}{\widetilde{m}_{\scriptscriptstyle{R}}}
\newcommand{\Tvac}{T_\textrm{vac}}
\newcommand{\WeylR}{{\cal W}_{\scriptscriptstyle{R}}}
\newcommand{\zeroL}{{0,\scriptscriptstyle{L}}}
\newcommand{\zeroR}{{0,\scriptscriptstyle{R}}}

\newcommand{\blue}{\textcolor{blue}}

\newcommand{\Prob}{{\rm Prob}}

\begin{document}

\title{ Last-passage time for linear diffusions and application to the  emptying time of a box
}


\author{\author{Alain Comtet \and Fran{\c c}oise Cornu \and Gr\'{e}gory Schehr} 
}


\institute{A. Comtet \at Universit\'e Paris-Saclay, CNRS, LPTMS, 91405, Orsay, France 
\\ \and F. Cornu \at Universit\'e Paris-Saclay, CNRS, LPTMS, 91405, Orsay, France \\
\and G. Schehr \at  Universit\'e Paris-Saclay, CNRS, LPTMS, 91405, Orsay, France}

\date{}

\maketitle

\begin{abstract}
We study the statistics of last-passage time for linear diffusions. First we present an elementary derivation of the Laplace transform of the probability density of the last-passage time, thus recovering known results from the mathematical literature. We then illustrate them on several explicit examples. In a second step we study the spectral properties of the Schr\"{o}dinger operator associated to such diffusions in an even potential $U(x) = U(-x)$, unveiling the role played by the so-called  Weyl coefficient. Indeed, in this case, our approach allows us to relate the last-passage times for dual diffusions (i.e., diffusions driven by opposite force fields) and to obtain new explicit formulae for the mean last-passage time. We further show that, for such even potentials, the small time $t$ expansion of the mean last-passage time on the interval $[0,t]$ involves the Korteveg-de Vries invariants, which are well known in the theory of Schr\"odinger operators. Finally, we apply these results to study the emptying time of a one-dimensional box, of size $L$, containing $N$ independent Brownian particles subjected to a constant drift. In the scaling limit where both $N \to \infty$ and $L \to \infty$, keeping the density $\rho = N/L$ fixed, we show that the limiting density of the emptying time is given by a Gumbel distribution. Our analysis provides a new example of the applications of extreme value statistics to out-of-equilibrium systems.
\end{abstract}







\section{Introduction}
First-passage times of random walks or diffusion processes have been extensively studied in the physics literature \cite{Hangi,Redner,BMS,BV2014}. Several physical or chemical properties  are controlled by first-passage events, a key example being reaction rates. In the Kramers approach they are given in terms of the  inverse mean first-passage time \cite{Hangi}. First-passage times also  play a crucial role in the study of transport properties, in particular in the context of biochemical reactions \cite{BV2014,Coppey,Holcman} 
and in the description of persistence properties of spatially extended systems \cite{BMS}.
However, there are several cases of physical interest where the mean first-passage does not give the proper time scale. In particular,  in nuclear physics, it is known that the fission rates of heavy nuclei are best described in terms of \textit{last-passage} time events \cite{Bao1}. The simpler case of the overdamped motion of a particle in a potential well already allows for a better understanding of the problem. In this case,  it has been  shown \cite{Bao2} that the mean last-passage time provides a more accurate formula for the escape rate at a saddle point than the usual mean first-passage time.

The purpose of this paper is to revisit the concept of last-passage time in the case of one-dimensional diffusion processes (see Fig. \ref{fig:FPT}) and illustrate its physical relevance by studying the emptying time of a box, namely the first time after which the box, containing initially a large number of particles, remains empty forever. The concept of last-passage time has been studied in detail in the mathematical literature. For instance, for Brownian motion, it is well known that the probability density of the last-passage at the origin is given by the famous ``arcsine law'' \cite{Levy,Feller}. More recently, last-passage times have also been studied in mathematical finance where last exit problems are commonly used, in particular as theoretical models of default risks \cite{Jeanblanc}.
In contrast, it seems that last-passage times have not been much exploited in the physics literature with the exception of Ref. \cite{Comdes} where  the last exit  distribution from a wedge is obtained in closed form as well as Ref.~\cite{Bendes} which deals with the last-passage problem on graphs. More recent works have also studied last-passage times for other processes like the fractional Brownian motion \cite{Kay} or for the run-and-tumble particle (also known as the persistent random walk model) \cite{Sing}. In this paper we provide an elementary derivation of the probability laws of the last-passage time in various cases and make contact with results known in the mathematical literature. We provide detailed formulae in terms of spectral properties of the generator of the diffusion which are very useful for explicit calculations. Similar spectral  representations, which have  been used in the study of the local and occupation time of a particle diffusing in a random medium \cite{Majcom,Sabha}, also appear in the so-called spectral theory of Krein strings \cite{Bertoin,Dym,Comtour}.

As an interesting application of these general formulae, we calculate the statistics of the emptying time of a box, $T_{\rm vac}$ (the subscript ``vac'' refereeing to ``vacuum''), for the following simple one-dimensional model: a box of size $L$  and opened on the right-hand side contains a gas consisting of $N$ independent particles, each of  which is reflected on the left wall of the box located at the origin and subjected to a white noise and  a constant drift towards the right (see Fig. \ref{Fig_vacuum} below).
In the  initial state the particles are uniformly distributed on the interval $[0,L]$. Due to the drift term, each particle $i=1,2, \cdots, N$ will eventually leave the box for ever, with probability one, at some random time $\tau^{(i)}$. This time coincides with {\it the last-passage time in $L$} for the $i$-th particle. Our aim is to compute the probability distribution of {the emptying time $T_{\rm vac}$} after which no particle will ever  again go into the box : thus  $T_{\rm vac}$  coincides with the largest last-passage times at position $L$ among the $N$ particles, i.e. $T_{\rm vac} = \max \{\tau^{(1)}, \tau^{(2)}, \cdots, \tau^{(N)}\}$. Note that since the ordering of the particles does not matter, the problem here is somewhat simpler than the one treated in \cite{Krap}. Thus we see that this emptying time problem can be formulated as an extreme value question for a set of $N$ independent {\it but non identically distributed} random variables, since the random variable $\tau^{(i)}$ depends on the initial position of the $i$-th particle. In the scaling limit where both $N$ and $L$ are large, at fixed density $\rho = N/L$, we show that the distribution of the emptying time, properly centered and scaled, converges to a Gumbel distribution [see Eq. \eqref{proba_Gumbel}], which is well known in the theory of extreme value statistics (EVS) for independent and identically distributed random variables \cite{Gum58} (for a recent review see \cite{MPS2020}).  
This result is interesting because it provides a new physical observable, as other ones found before like the convex hull of $N$ planar Brownian motions \cite{Randon,Randon_2} or the number of distinct and common sites visited by $N$ independent random walkers \cite{Kundu}, which is directly related to extreme value statistics. In fact the connection between exit-time problems and extreme value statistics goes back to Ref. \cite{Day} and it was recently extended in Ref. \cite{Bak1} and used to model the incubation period of certain diseases \cite{Bak2}. In these works only the first exit time is considered and the connection with extreme value statistics only holds in the limit of vanishing noise. Our work is in some sense complementary since we focus here on the last exit time and explore different scaling regimes.  

The rest of the paper is organized as follows. In Section II, we present a general framework to compute the distribution of the last-passage time $g_{x_\star,t}$ at position $x_\star$ up to the observation time $t$, for a general linear diffusion starting from $x_0$. This leads to a rather explicit expression for the double Laplace transform of the distribution of $g_{x_\star,t}$ with respect to both its argument as well as the observation time $t$, given in Eq. (\ref{DoubleLTBis}). In Section III, we apply this general formalism to several specific cases, where $x_0=x_\star$,  
for which explicit formulae can be obtained: this includes in particular the Brownian motion with a drift \eqref{formula_BM_drift}, the hyperbolic tangent force (\ref{Probtanh}), the Ornstein-Uhlenbeck process \eqref{ProbOU} as well as the Bessel process reflected at the origin \eqref{Prob_Bessel}. Section~IV is devoted to the case of a force deriving from an even potential $U(x) = U(-x)$ (again with $x_0=x_\star$). In this case, we demonstrate the key role played by the so-called ``Weyl coefficient'' [see Eqs. \eqref{defWeyl} and \eqref{LTmeanValueUevenBis}], which plays an important role in the theory of Sturm-Liouville operators. We obtain an explicit expression for the average last-passage time in terms of quantum partition functions (\ref{RateMeanValue}) associated to the Schr\"odinger operator corresponding to the diffusion. This, in turn, allows us to obtain the short time $t$ expansion of the average last-passage time up to time $t$ in terms of the Korteweg-de Vries (KdV) invariants [see Eq. \eqref{RateMeanValue3}]. In Section V, we consider the limit of an infinite observation time $t \to \infty$ for a transient process. In this case, we obtain fairly explicit expressions for different transient behaviours (at $+\infty$ and/or $-\infty$) and various boundary conditions, as given in Eqs.~\eqref{LimInfiniteDurationCase1}-\eqref{LimInfiniteDuration4}. In Section VI, we apply the results of Section V to the specific example of the reflected Brownian motion with a constant drift, for which we provide an interesting semi-classical interpretation, together with an explicit expression of the density of the last-passage time in the limit of an infinite observation time [see Eqs. \eqref{propa} and \eqref{densityprobagdriftrefl}]. In Section~VII, we use these results on last-passage times to compute the distribution of the emptying time $T_{\rm vac}$ of a box of size $L$, for $N$ independent particles performing a reflected Brownian motion with a constant drift, as mentioned above. Finally, Section~VIII contains our conclusion and perspectives. Some technical details have been relegated to Appendices A to~E.

\section{Last-passage time distribution: general framework}

\subsection{Linear diffusions on the real axis}

We consider the time-homogeneous  one-dimensional overdamped Langevin equation
\be
\label{LangevinEq}
\dd x(t)= F(x)\, \dd t + \sqrt{2D}\,\dd B(t),
\ee
where 
\be
F(x)=-\frac{\dd U(x)}{\dd x}
\ee 
is a time-independent  force and $B(t)$ is a Brownian motion   $\dd B(t)= \eta(t) \dd t$ where $\eta(t)$ is a  Gaussian white noise with zero mean $\mathbb{E}\left[ \eta(t)\right]=0$  and variance $\mathbb{E}\left[ \eta(t)\eta(t') \right]=\delta(t-t')$. In Eq. (\ref{LangevinEq}), $D$ is the diffusion constant.

The central object is the transition kernel $P(x,t \vert x_0, 0)$ for paths which start at $x_0$ at time $0$ and go through $x$ at time $t$. The transition kernel obeys the forward Fokker-Planck equation
\be
\label{ForwardFK}
\frac{\partial }{\partial t} P(x,t \vert x_0, 0) = \Gen^\dag_x P(x,t \vert x_0, 0) \;,
\ee
where 
\be
 \Gen^\dag_x P(x,t \vert x_0, 0)= D \frac{\partial^2}{\partial x^2}  P(x,t \vert x_0, 0)- \frac{\partial}{\partial x} \left[ F(x) P(x,t \vert x_0, 0) \right].
\ee
Here $\Gen^\dag_x$ is the adjoint of the diffusion operator which governs the backward Fokker-Planck equation
\be
\label{BackwardFK}
\frac{\partial }{\partial t} P(x,t \vert x_0, 0) = \Gen_{x_0}P(x,t \vert x_0, 0),
\ee
where
\be
\label{Generator}
\Gen_x =D \frac{\partial^2}{\partial x^2} + F(x) \frac{\partial }{\partial x}.
\ee
We denote by $T_{x_\star}$ the first-passage time at position $x_\star$
\be
T_{x_\star}=\inf\left\{  \tau \geq 0, x(\tau)=x_\star\right\}
\ee
and  $g_{x_\star,t}$ the last-passage time at position $x_\star$ during the observation time $t$,
\be
g_{x_\star,t}=\sup\left\{  \tau\leq t, x(\tau)=x_\star\right\}.
\ee
{In the following, $x_\star$ will be sometimes called the ``target position'' or ``target level''}. 
Our primary focus is to compute the probability distribution of $g_{x_\star,t}$ for linear diffusion processes described by the Langevin equation~(\ref{LangevinEq}). Before that, we briefly recall some well known results for the first-passage distribution.

\subsection{A reminder on the distribution of the first-passage time}
\label{sec:FirstPassageTime}

\begin{figure}[h]
\begin{center}
\includegraphics[width=0.7\linewidth]{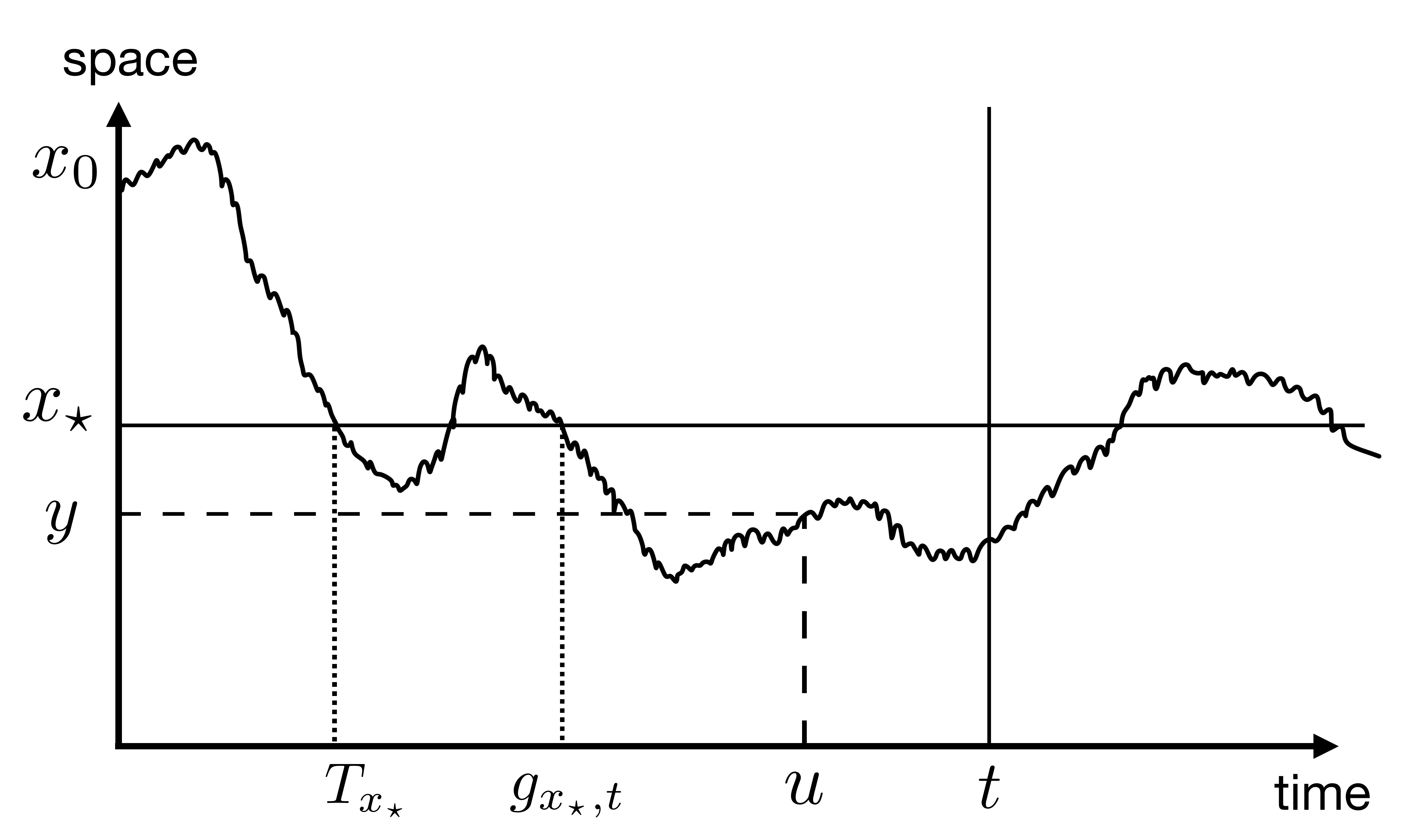}
\end{center}
\caption{{Case $x_\star<x_0$: sketch of a path starting at $x_0$ and arriving at $y$ at a given time $u \in [0,t]$ such that $y<x_\star$. The time $T_{x_\star}$ denotes the first-passage time at $x_\star$ while $g_{x_\star,t}$ denotes
the last-passage time at $x_\star$ on the time interval $[0,t]$. Such a path which starts at  $x_0$ and goes through $y$ at time $u$ necessarily also goes through $x_\star$ before $u$. Hence it contributes both to $\Prob_{x_0}(T_{x_\star} \leq u)$ and $\Prob_{x_0}(g_{x_\star,t}\leq u)$.}}
\label{fig:FPT}
\end{figure}

{In the case $x_\star<x_0$, let us consider all the paths starting at $x_0$, i.e. $x(0)=x_0$, and arriving at $y$ at a given time $u \in [0,t]$, i.e. $x(u) = y$, such that $y<x_\star$ (see Fig.~\ref{fig:FPT}).} These paths cross the level $x_\star$  
 and can thus be decomposed into two parts: (i) a first part that starts from $x(0)=x_0$ and hits level $x_\star$ for the first time at time $T_{x\star}$ and (ii) a second part which goes from $x_\star$ to $y$. For paths starting at $x_0$, we denote by $\Prob_{x_0}(T_{x_\star}<\tau)$ the probability that the first passage time $T_{x_\star}$ at $x_\star$ is less than some given time $\tau$. From the above path decomposition for  $y<x_\star$, the transition kernel can  be written as
\be
P(y,u \vert x_0,0) = \int_0^u P(y,u \vert x_\star,\tau) \, \frac{\partial \Prob_{x_0}(T_{x_\star} \leq \tau)}{\partial \tau} \, \dd \tau,
\ee
where $\partial  \Prob_{x_0}(T_{x_\star} \leq \tau)/ \partial \tau$  is the  first passage time density at $\tau$.
Since the process is time-homogeneous, i.e. translationally invariant in time, 
one has $P(y,u\vert x_\star,\tau)= P(y,u-\tau\vert x_\star,0)$ and the above integral is therefore a convolution. A  Laplace transform with respect to $t$  then gives
\be
\label{EspTexpression1}
\mathbb{E}_{x_0}\left[ \ed^{-\lambda T_{x_\star}}\right]=\int_0^{\infty} \ed^{-\lambda \tau}\,
\frac{\partial }{\partial  \tau}\Prob_{x_0}(T_{x_\star} \leq \tau)=\frac{\widehat{P}_\lambda(y\vert x_0)}{\widehat{P}_\lambda(y\vert x_\star)},
\ee
where $\mathbb{E}_{x_0}[\ldots]$ denotes an average all the trajectories starting at $x_0$ and we have used the notation
\be
\label{defwidehatP}
\widehat{P}_\lambda(y\vert x) = \int_0^{\infty} \ed^{-\lambda u}\, P(y,u\vert x,0)\, \dd u \;,
\ee
for the Laplace transform of the transition kernel with respect to the final time. As recalled in Appendix \ref{KernelExpression}, $\widehat{P}_\lambda(y\vert x)$ can be written in the following form
\be
\label{LTTransitionKernel}
\widehat{P}_\lambda(y\vert x)=\frac{1}{D}\frac{\phi_\lambdaL(x\land y) \phi_\lambdaR(x\lor y)}
{ \phi_\lambdaR(y) \phi'_\lambdaL(y) - \phi'_\lambdaR(y) \phi_\lambdaL(y)} \;,
\ee
where  $x\land y= \inf(x,y)$ (respectively $x\lor y=\sup(x,y)$) denotes the smaller (respectively the larger) one among $x$ and $y$, while $\phi_\lambdaR$ and $\phi_\lambdaL$ are the so-called fundamental solutions \cite{BORO} of the differential equation
\be
\label{Generateur}
\Gen_x\phi_\lambda= \lambda \phi_\lambda.
\ee
The functions $\phi_\lambdaR$ and $\phi_\lambdaL$, defined up to a multiplicative constant, are  such that $\phi_\lambdaR$ is non-increasing and obeys some right spatial boundary condition,   while $\phi_\lambdaL$ is non-decreasing and obeys some  left spatial boundary condition. The specific boundary conditions  obeyed by $\phi_\lambdaR$ and $\phi_\lambdaL$ are derived  from those obeyed by the transition kernel, as discussed  in Appendix A.  In the case of a diffusion along the whole real axis $\lim_{x\to +\infty} \ed^{-U(x)/D}\phi_\lambdaR(x)=0$ while  $\lim_{x\to -\infty} \ed^{-U(x)/D}\phi_\lambdaL(x)=0$. In the case of a diffusion along the positive real half-axis with reflection at $x=0$ the derivative of $\phi_\lambdaL$ vanishes at  $x=0$, $\phi'_{\lambdaL}(0)=0$.

The general expression for the transition kernel (\ref{LTTransitionKernel}), once inserted in Eq. (\ref{EspTexpression1}), yields a formula for the Laplace transform of the distribution of the first passage $T_{x^*}$. For all $y<x_\star<x_0$, $\inf(y,x_0)=\inf(y,x_\star)=y$  and after inserting the expression \eqref{LTTransitionKernel} into the ratio \eqref{EspTexpression1} we get
\be
\label{EspTx0supxstar}
\mathbb{E}_{x_0}\left[ \ed^{-\lambda T_{x_\star}}\right]=\frac{\phi_\lambdaR(x_0)}{\phi_\lambdaR(x_\star)}
\quad \textrm{if $x_0>x_\star$}.
\ee
In the case $x_0 <x_\star$ if we consider all paths such that $x(0)=x_0$ and $x(t)=y$ with $y<x_\star$ then a similar argument leads to
\be
\label{EspTx0infxstar}
\mathbb{E}_{x_0}\left[ \ed^{-\lambda T_{x_\star}}\right]=\frac{\phi_\lambdaL(x_0)}{\phi_\lambdaL(x_\star)}
\quad \textrm{if $x_0<x_\star$}.
\ee
As we will see, these expressions (\ref{EspTx0infxstar}) and (\ref{EspTx0supxstar}) will be useful for the computation of the probability density of the last-passage time, which we now focus on.

\subsection{Last-passage time}
\subsubsection{Last-passage time density}


\begin{figure}[h]
\begin{center}
\includegraphics[width = 0.7\linewidth]{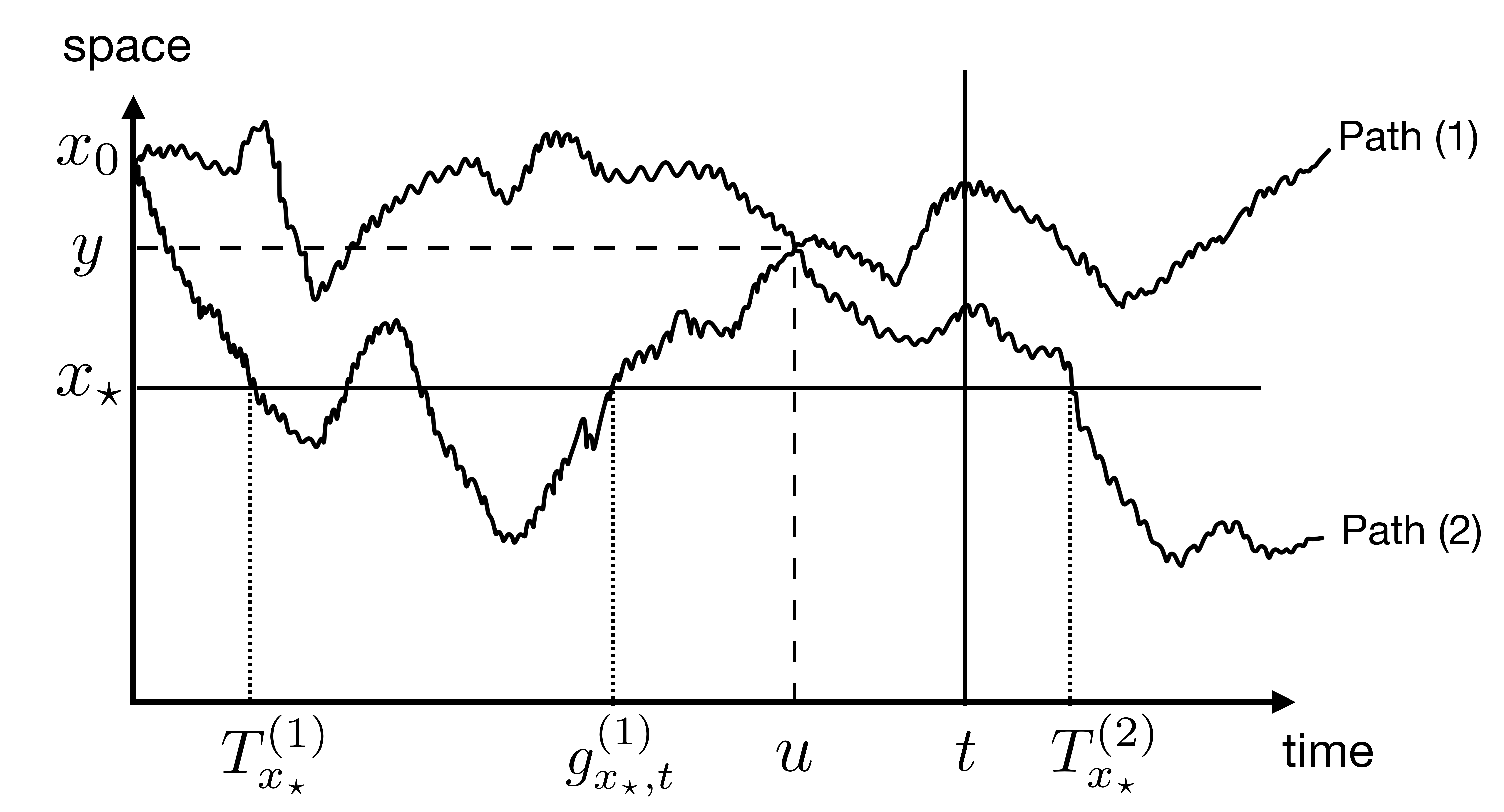}
\end{center}
\caption{\small {Case $x_\star < x_0$: sketch of two paths starting at $x_0$ and arriving at $y$ at a given time $u \in [0,t]$ such that $y>x_\star$. Such paths do not necessarily go through $x_\star$ before $u$.  For example, path (1) contributes both to $\Prob_{x_0}(T_{x_\star} \leq u)$ and $\Prob_{x_0}(g_{x_\star,t}\leq u)$, but  path (2) does not go through $x_\star$ in the time interval $[0,t]$ (the crossing happens later):  therefore path (2)  does not contribute to  $\Prob_{x_0}(T_{x_\star}\leq u)$ but it contributes to the event ``$g_{x_\star, t} = 0$''  and hence to $\Prob_{x_0}(g_{x_\star,t}\leq u)$ because it never crosses $x_\star$ after time $0$ in the observation window $[0,t]$.}}\label{fig:LPT}
\end{figure}

In the case of the last-passage time before the observation time $t$  the argument is slightly more subtle than for the first-passage time. It requires a careful analysis of  all paths such that $x(0)=x_0$ and $x(t)$ takes any value. For each path one can define the random variable $g_{x_\star,t}$ as the last time during the interval $[0,t]$ at which the path crosses  level $x_\star$ (see Fig. \ref{fig:FPT}). The last-passage time can also be considered as the time after which the path never crosses the level $x_\star$ before the last observation time $t$; as a consequence $g_{x_\star,t}=0$ if the path never crosses $x_\star$ during the time interval $[0,t]$. Then for a given $u<t$, a path for which $g_{x_\star,t}\leq u$ can be decomposed into two parts~(see Fig. \ref{fig:LPT}): the first part  goes from $x_0$ to some value $y$ (and possibly never crosses $x_\star$) during the time interval $[0,u]$, while the second part starts from $y$ and never crosses $x_\star$ during the remaining time interval $[u,t] $ (and goes through  any arbitrary level  at time $t$). The probability weight associated to the first part of the path is the transition kernel $P(y,u \vert x_0,0)$, and the weight of the second part is  $\Prob_y(T_{x_\star}>t-u)$. Eventually  the cumulative distribution function of the last-passage time reads
 \bea
\label{Cumulative}
\Prob_{x_0}(g_{x_\star,t}\leq u)&=&
\int_{x_\star}^{+\infty} \,\Prob_y\left( T_{x_\star} > t-u \right) P(y, u \vert x_0, 0) \,\dd y \nonumber \\
&+&\int_{-\infty}^{x_\star}  \,\,\Prob_y\left( T_{x_\star} > t-u \right)P (y, u \vert x_0, 0) \, \dd y \;.
\eea
We have split the integral into two parts  ($y>x_\star$ and $y<x_\star$) 
because the expression for the probability $\Prob_y(T_{x_\star}>t-u)$ depends on the relative positions of $y$ and $x_\star$, as can be checked from the expressions \eqref{EspTx0supxstar} and \eqref{EspTx0infxstar} for the  Laplace transforms of the first-passage time density 
for each relative position.

It is convenient to introduce the last-passage time density $\Pi_{g_{x_\star,t}}(u)$, whose integral is equal to the cumulative distribution function : $\Prob_{x_0}(g_{x_\star,t}\leq u)=\int_0^u \Pi_{g_{x_\star,t}}(u)\, \dd u$.
This density is the sum of a smooth part, the probability that $g_{x_\star,t}$ lies in $]u, u+ \dd u]$, namely $\frac{\partial \Prob_{x_0}(g_{x_\star,t}\leq u)}{\partial u} \dd u$, and a Dirac mass at $u=0$
\be
\label{decomPi}
\Pi_{g_{x_\star,t}}(u) = \Prob_{x_0}(T_{x_\star}> t) \, \delta(u) + \frac{\partial \Prob_{x_0}(g_{x_\star,t}\leq u)}{\partial u}.
\ee
The Dirac mass comes from those paths which never cross the level $x_\star$ during $[0,t]$. Their weight is equal to the probability that the first passage time at $x_\star$ is larger than $t$. Indeed by letting $u\rightarrow 0^{+}$ in \eqref{Cumulative} we get
\be
\Prob_{x_0}(g_{x_{\star},t}=0)=\Prob_{x_0}(T_{x_\star}> t).
\ee
In the following we shall refer to the smooth part as the last-passage time ``density''. Note that, as a consequence of  \eqref{decomPi}, the normalization of the  latter density reads
\be
\label{NormalizationLastPassageDensity}
\int_0^t\frac{\partial \Prob_{x_0}(g_{x_\star,t}\leq u)}{\partial u}\, \dd u= \Prob_{x_0}(T_{x_\star}\leq t).
\ee
The density is said to be ``defective'' (as it is not normalized to one) and the identity \eqref{NormalizationLastPassageDensity} will be checked explicitly below in the case of the Brownian motion.

We aim to derive an expression of the density starting from \eqref{Cumulative}. 
The partial derivative with respect to $u$ involves  both the first-passage time probability and the transition kernel.
The equation satisfied by the 
the first-passage time probability for $y>x_\star$ and $y<x_\star$ can be derived from the expression of the survival probabilities in terms of two independent Dirichlet Green's functions, 
\bea
\label{relProbTxstarGDxstar}
w_{ \scriptscriptstyle{R}}(y,t-u)=\Prob_y\left( T_{x_\star} > t-u \right)&=& \int_{x_\star}^{+\infty} G_{\textrm{D},x_\star} (x,t \vert y,u) \,\dd x \qquad \textrm{if $y>x_\star$} \;,
\\
\nonumber
w_{ \scriptscriptstyle{L}}(y,t-u)= \Prob_y\left( T_{x_\star} > t-u \right)&=& \int_{-\infty}^{x_\star} \,\, G_{\textrm{D},x_\star} (x,t \vert y,u) \,\dd x \qquad \textrm{if $y<x_\star$} \;.
\eea
From this representation, it follows that $w_{\scriptscriptstyle{R,L}}(y,t-u)$ obey the backward Fokker-Planck equation 
\be
\frac{\partial}{\partial u} w(y,t-u) =- \Gen_y w(y,t-u) \;,
\ee
with the boundary conditions $w_{ \scriptscriptstyle{R}}(y,0)=1$ for $ y>x_\star$,  $w_{ \scriptscriptstyle{L}}(y,0)=1$  for $ y<x_\star$ and $\forall t>0, w_{ \scriptscriptstyle{R}} (x^{\star},t)= w_{ \scriptscriptstyle{L}} (x^{\star},t)=~0 $. On the other hand, the transition kernel $P (y, u \vert x_0, 0)$ obeys the forward Fokker-Planck equation [see Eq.~\eqref{ForwardFK}]. Then, remarkably, one can check that the derivative with respect to $u$ of each integral in \eqref{Cumulative} can be rewritten as the integral over $y$ of a total derivative with respect to $y$. Thus  the integration over $y$ can be performed explicitly and gives two boundary terms,
\be
\frac{\partial}{\partial u} \Prob_{x_0}(g_{x_\star,t}\leq u)= \left[ K(y) \right]_{-\infty}^{x_\star^-} + \left[ K(y) \right]_{x_\star^+}^{+\infty}
\ee
with
\be
K(y)=-D P (y, u \vert x_0, 0) \frac{\partial}{ \partial y} \Prob_y\left( T_{x_\star} > t-u \right)
+ \Prob_y\left( T_{x_\star} > t-u \right) \left[ D \frac{\partial}{\partial y} + \frac{\dd U}{\dd y}\right] P (y, u \vert x_0, 0).
\ee
We notice that on the r.h.s. the second term   coincides with the probability current at $y$ at time $u$, $J_P(y,u\vert x_0,0)$, by virtue of  \eqref{DefJP}. When $y\to +\infty$ and $y\to -\infty$ all probabilities vanish as well as  the probability current. In the case where the boundary $-\infty$ is replaced by a reflecting boundary at $x=0$ the probability current 
$J_P(y,u\vert x_0,0)$  vanishes as well as $\partial \Prob_y\left( T_{x_\star} > t-u \right)/\partial y\vert_{y=0}$, as shown in Appendix \ref{FormulaIntermediate}.
Eventually, for any boundary condition of interest,  the last-passage time density is given by the two boundary terms at $x_\star^-$ and $x_\star^+$ and it factorizes into a function of $t-u$ times a function of $u$
\be
\label{Product}
\frac{\partial \Prob_{x_0}(g_{x_\star,t}\leq u)}{\partial u}=  h_{t-u}(x_\star)\, P (x_\star, u \vert x_0, 0) \;,
\ee
where the ``boundary term'' $h_\tau(x_\star)$ reads
\be
\label{defh}
h_\tau(x_\star)=D \left[ 
\left. \frac{\partial}{\partial y}\Prob_y \left(T_{x_\star}>\tau \right)  \right\vert_{y=x_\star^+}
- \left. \frac{\partial}{\partial y}\Prob_y \left(T_{x_\star}>\tau \right)  \right\vert_{y=x_\star^-}
\right]. 
\ee
One can provide a heuristic interpretation of the formula for the density in Eq. (\ref{Product}). Roughly speaking, the probability density that
the last-passage in $x_{\star}$ is $u$, i.e. $\partial \Prob_{x_0}(g_{x_\star,t}\leq u)/\partial u $, is the probability to go through $x_{\star}$ at time $u$, which is given by $P (x_\star, u \vert x_0, 0)$, 
multiplied by the probability for a trajectory starting from $x_\star$ not to come back to $x_\star$ during the time interval $(t-u)$.
This interpretation is made clearer if the final point $x(t)=x$, is kept fixed. Then the trajectories which contribute  to the density can be decomposed into two parts:
\begin{itemize}
\item[$\bullet$] A bridge, i.e. a path $x(\tau) $ such that  $ x(0)= x_{0}$ and $ x(u) = x_{\star}$;
\item[$\bullet$] A meander, i.e. a path $x(\tau)$ such that  $x(u)= x_{\star}$, $ x(t) =x $ with the constraint that $x(\tau)>x_{\star} $ $\forall \tau\in] u,t ]$ (upper meander) or $x(\tau) < x_{\star}$ $\forall \tau\in] u,t ]$ (lower meander). 
\end {itemize}
Then one can prove that the two boundary terms in Eq. (\ref {defh}) come respectively from the contribution of the upper and lower meanders. 

\subsubsection{Double Laplace transform}

It turns out to be useful to consider the double Laplace transform of the last-passage time density with respect to the two time variables $t$ and $u$, with $0<u<t$,
\be
\label{defDoubleLT}
\int_0^{+\infty} \ed^{-\lambda t} \, \mathbb{E}_{x_0}\left[\ed^{-\lambda' g_{x_\star, t}} \right]\, \dd t=
\int_0^{+\infty} \ed^{-\lambda t} \, \dd t \int_0^t \ed^{-\lambda' u} \, \frac{\partial \Prob_{x_0}(g_{x_\star,t}\leq u)}{\partial u}\, \dd u \;.
\ee
According to the factorization property \eqref{Product} the double  Laplace transform is also a product
\be
\label{DoubleLT}
\int_0^{+\infty} \ed^{-\lambda t} \, \mathbb{E}_{x_0}\left[\ed^{-\lambda' g_{x_\star, t}} \right]\, \dd t=
\widehat{h}_{\lambda}(x_\star) \widehat{P}_{\lambda+\lambda'} (x_\star \vert x_0) \;,
\ee
where we have used the same notation for the Laplace transform as in \eqref{defwidehatP} :
$\widehat{h}_\lambda(x_\star) = \int_0^{\infty} \ed^{-\lambda \tau}\,h_\tau(x_\star) \,\dd \tau$.

As in the case of the first-passage time, it is convenient to rewrite everything in terms of the fundamental solutions of the diffusion generator $\Gen_x$ and their derivatives. This standard approach in the study of linear diffusion processes \cite{BORO} has been also used in the physics literature \cite{Majcom,Sabha}. The Laplace transform $\widehat{h}_\lambda(x_\star) $ can be obtained by inserting the expression \eqref{relLT} derived in Appendix \ref{FormulaIntermediate}
into the definition \eqref{defh} and then, using the expressions \eqref{EspTx0infxstar} and \eqref{EspTx0supxstar} for $ \mathbb{E}_{x_0}\left[\ed^{-T_{x_\star}} \right]$, we obtain the final result
\be
\label{LTh}
 \widehat{h}_{\lambda}(x_\star) = - D   \frac{1}{\lambda} \left[ 
\left.\frac{ \phi'_\lambdaR(y)}{\phi_\lambdaR(y)}\right\vert_{y=x_\star}
 - \left.\frac{ \phi'_\lambdaL(y)}{\phi_\lambdaL(y)}\right\vert_{y=x_\star}
 \right] \;,
\ee
where $\phi'_{\lambda, R/L}(y)$ denotes the derivative of $\phi_{\lambda, R/L}(y)$ with respect to $y$. 
The latter expression involves logarithmic derivatives which will play a crucial role in the following
\be
\label{mRmLdefinition}
\mR(\lambda; x_\star) = \left.\frac{ \phi'_\lambdaR(y)}{\phi_\lambdaR(y)}\right\vert_{y=x_\star}
\qquad, \qquad 
\mL(\lambda; x_\star)=\left.\frac{ \phi'_\lambdaL(y)}{\phi_\lambdaL(y)}\right\vert_{y=x_\star},
\ee
and it can be rewritten as
\be
\label{LThBis}
 \widehat{h}_{\lambda}(x_\star) =  D  \frac{ \mL(\lambda; x_\star)- \mR(\lambda; x_\star)}{\lambda}  \;.
\ee
For the Laplace transform of the transition kernel,  comparison of its expression \eqref{LTTransitionKernel} in terms of the eigenfunctions of the diffusion generator with the corresponding expressions \eqref{EspTx0supxstar} and \eqref{EspTx0infxstar} for $ \mathbb{E}_{x_0}\left[\ed^{-T_{x_\star}}\right] $ leads~to 
\be
\label{LTTransitionKernelBis}
\widehat{P}_\lambda (x_\star \vert x_0)= \frac{1}{D}
\frac{1}{\mL(\lambda; x_\star)- \mR(\lambda; x_\star)} \, \mathbb{E}_{x_0}\left[\ed^{-\lambda T_{x_\star}} \right] \;.
\ee
By inserting \eqref{LThBis} and \eqref{LTTransitionKernelBis} into \eqref{DoubleLT} we get
\be
\label{DoubleLTBis}
\int_0^{+\infty} \ed^{-\lambda t} \, \mathbb{E}_{x_0}\left[\ed^{-\lambda' g_{x_\star, t}} \right]\, \dd t=
\frac{1}{\lambda}
\frac{m(\lambda ; x_\star)}{m(\lambda+\lambda' ; x_\star)}\, \mathbb{E}_{x_0}\left[\ed^{-(\lambda +\lambda' )T_{x_\star} }\right] \;,
\ee
where the expectation  $\mathbb{E}_{x_0}\left[\ed^{-\lambda' g_{x_\star, t}} \right]$ is to be taken only with respect to the smooth part of the last-passage time density, see \eqref{defDoubleLT}, and where  we have introduced the notation
\be 
\label{mdefinition}
m(\lambda ; x_\star)=\mL(\lambda ; x_\star)-\mR(\lambda ; x_\star) \;.
\ee

In the case where the target level $x_\star$ coincides with the initial value $x_0$, $T_{x_\star} = 0$ and therefore  $\mathbb{E}_{x_\star}\left[\ed^{-\lambda T_{x_\star}} \right]~=~1$ for all $\lambda \geq 0$. Then the last passage time density \eqref{decomPi} is reduced to its smooth part. Therefore the double Laplace transform \eqref{DoubleLTBis} takes the simple form
\be
\int_0^{+\infty} \ed^{-\lambda t} \, \mathbb{E}_{x_\star}\left[\ed^{-\lambda' g_{x_\star, t}} \right]\, \dd t=\frac{1}{\lambda}
\frac{m(\lambda ; x_\star) }{m(\lambda+\lambda' ; x_\star) } \;.
\ee
One can check that the latter formula is equivalent to the expression given on page 27  in \cite{BORO}   in terms of the so-called Green's function associated to the diffusion generator. A proof in the context of subordinators  is given in \cite{Bertoin}.
Finally, we notice that  if the  potential $U(x)$ is even, i.e. $U(x)=U(-x)$, one can show that $\phi_\lambdaR(x)=\phi_\lambdaL(-x)$. Then $\mR(\lambda; x_\star)=-\mL(\lambda; x_\star)$ and
\be
\label{DoubleLTexpression}
\int_0^{+\infty} \ed^{-\lambda t} \, \mathbb{E}_{x_\star}\left[\ed^{-\lambda' g_{x_\star, t}} \right]\, \dd t=\frac{1}{\lambda}
\frac{\mR(\lambda; x_\star)}{\mR(\lambda+\lambda' ; x_\star)} \;,
\ee
which will be useful in the following.

\subsubsection{Mean last-passage time}

The Laplace transform of the mean last-passage time with respect to the observation time can be obtained from the double Laplace transform given in  \eqref{DoubleLTBis}. It reads
\be
\label{MeanValue}
\int_0^{+\infty} \ed^{-\lambda t} \, \mathbb{E}_{x_0}\left[g_{x_\star, t} \right]\, \dd t=
\frac{1}{\lambda}\left(
\frac{\dd  \ln m(\lambda ; x_\star) }{\dd \lambda}{\mathbb{E}_{x_0}\left[\ed^{-\lambda T_{x_\star}} \right]}- \frac{\dd }{\dd \lambda}\mathbb{E}_{x_0}\left[\ed^{-\lambda T_{x_\star}} \right]
\right)
\ee
where  the expression of $\mathbb{E}_{x_0}\left[\ed^{-\lambda T_{x_\star}} \right]$ in terms of the  fundamental solutions
$\phi_\lambdaR$ and $\phi_\lambdaL$  is given in \eqref{EspTx0infxstar} and \eqref{EspTx0supxstar} while $m(\lambda;x_\star)$ is given in \eqref{mdefinition}. The mean last-passage time can thus be obtained by inverse Laplace transform of the latter relation~\eqref{MeanValue}.

\subsubsection{Explicit formulae for the Brownian  motion} 

In the case of a one-dimensional Brownian motion, described by Eq. (\ref{LangevinEq}) with $F(x)=0$, the generator~\eqref{Generator} reads
$\Gen_x=D \partial^2/\partial x^2$
and the  corresponding solutions defined in \eqref{Generateur} are simply given~by
\be
\label{eigenfBrownian}
\phi_\lambdaR(x)= \ed^{-x \sqrt{\lambda/D}}
\qquad\textrm{and}\qquad
\phi_\lambdaL(x)= \ed^{x \sqrt{\lambda/D}}.
\ee
The Laplace transform of the transition kernel, $\widehat{P}_\lambda(y\vert x)$, can be obtained from its representation \eqref{LTTransitionKernel} :  $\widehat{P}_\lambda(y\vert x)= 1/(2\sqrt{\lambda D}) \ed^{-\sqrt{\lambda/D}\vert x-y\vert}$. By inverse Laplace transform of the definition \eqref{defwidehatP} we recover the well-known kernel
\be
\label{kernelBrownian}
P(y,t\vert x,0)=\frac{1}{\sqrt{4 \pi D t}}\ed^{-(x-y)^2/4Dt} \;.
\ee
The first-passage time density can also be determined explicitly by Laplace inversion of the representations
\eqref{EspTx0supxstar}-\eqref{EspTx0infxstar}, $\mathbb{E}_{x_0}\left[ \ed^{-\lambda T_{x_\star}}\right]=\ed^{-\sqrt{\lambda/D}\vert x_0-x_\star\vert}$, and we retrieve the well-known first-passage time distribution for a Brownian motion
\be
\frac{\partial \Prob_{x_0}(T_{x_\star}\leq \tau)}{\partial \tau}=\frac{1}{\sqrt{4 \pi D }} \frac{\vert x_\star-x_0 \vert}{\tau^{3/2}}\ed^{-( x_\star-x_0)^2/4D\tau} \;.
\ee
The cumulative distribution for the first-passage time is readily  obtained by integration over $\tau$ with the result
\be
\label{probfirstpassageBrownian}
\Prob_{x_0}(T_{x_\star}\leq \tau)= 1-\erf\left(\frac{\vert x_\star-x_0 \vert }{\sqrt{4D \tau}} \right) \;,
\ee
where the error function reads
\be
\label{erfdef}
\erf(\tau) = \frac{2}{\sqrt{\pi}} \int_0^\tau  \ed^{-u^2} \dd u \;.
\ee

The last-passage time density can be  obtained from the product representation \eqref{Product} : the expression for the  kernel  has been given in \eqref{kernelBrownian}  and the function $h_t(x_\star)$  can be determined from its Laplace transform given in \eqref{LTh}. The expressions of the eigenfunctions \eqref{eigenfBrownian} lead to
$\widehat{h}_\lambda(x_\star)=2 \sqrt{D/\lambda}$, and after inverse Laplace transform we get 
\be
h_t(x_\star)=\sqrt{\frac{4D}{\pi}}\frac{1}{\sqrt{t}} \;.
\ee
 Eventually the last-passage time probability density reads
\be
\label{lastBrown}
\frac{\partial \Prob_{x_0}(g_{x_\star,t} \leq u)}{\partial u}=
 \frac{1}{\pi\sqrt{u(t-u)}} \ed^{-(x_\star-x_0)^2/4Du} \;.
\ee
In the special case $x_\star = x_0$, this expression yields back the celebrated arcsine law [see also Eq. (\ref{ProbBrownian00}) below]. By integration over $u$ we get
\be
\int_0^t\frac{\partial \Prob_{x_0}(g_{x_\star,t} \leq u)}{\partial u} \, \dd u= 1 -\erf\left(\frac{\vert x_\star-x_0 \vert }{\sqrt{4D t}} \right) \;.
\ee
By comparison with the expression \eqref{probfirstpassageBrownian} for $\Prob_{x_0}(T_{x_\star}\leq \tau)$ we check that the normalization \eqref{NormalizationLastPassageDensity} is indeed satisfied.
The mean value of the last-passage time can be obtained by Laplace inversion of the relation \eqref{MeanValue}, using 
the definition of $m(\lambda ; x_\star)$ in \eqref{mdefinition}, which implies that  $\dd \ln m(\lambda;x_\star)/\dd \lambda= 1/(2\lambda)$, while $\mathbb{E}_{x_0}\left[\ed^{-\lambda T_{x_\star}} \right] = \ed^{-|x_\star - x_0|\sqrt{\lambda/D}}$ from \eqref{EspTx0infxstar}, \eqref{EspTx0supxstar} and \eqref{eigenfBrownian}. This yields

\bea
\label{meangBrownian}
 \mathbb{E}_{x_0}\left[g_{x_\star, t} \right]
 &=& t \, \overline{g}\left(\frac{|x_\star-x_0|}{\sqrt{4 D \,t}} \right) \;, \; \bar{g}(u) = \left(\frac{1}{2}-u^2 \right){\rm Erfc(u)} + \frac{u}{\sqrt{\pi}}\ed^{-u^2}\;,
 \eea
 where the complementary error function is defined as $\erfc(x)=1-\erf(x)$. Formula \eqref{meangBrownian} can also be obtained directly from the probability density in \eqref{lastBrown}. For $u \to 0$, one easily checks that $\bar{g}(u) \to 1/2$ while $\bar{g}(u) \sim \ed^{-u^2}/(\sqrt{\pi} u)$ as $u \to \infty$ (we recall that $g_{x_\star,t} = 0$ if the path never crosses the level $x_\star$).

\section{Solvable linear diffusion processes in the case $x_0=x_\star=0$}

In this section, we illustrate the general framework developed in the previous section, in particular the formula in Eq.~\eqref{DoubleLTexpression}, by considering a few solvable models in the case where  the initial and target points coincide with the origin, i.e. $x_0=x_\star=0$.

\subsection{Brownian motion with constant drift}

In the case of a one-dimensional Brownian process with constant drift, $F(x)=\mu$, the generator \eqref{Generator} reads
\be
\label{Gendrift}
\Gen_x=D \frac{ \partial^2}{\partial x^2} + \mu \frac{\partial}{\partial x}.
\ee
\begin{figure}
\centering
\includegraphics[width = 0.7 \linewidth]{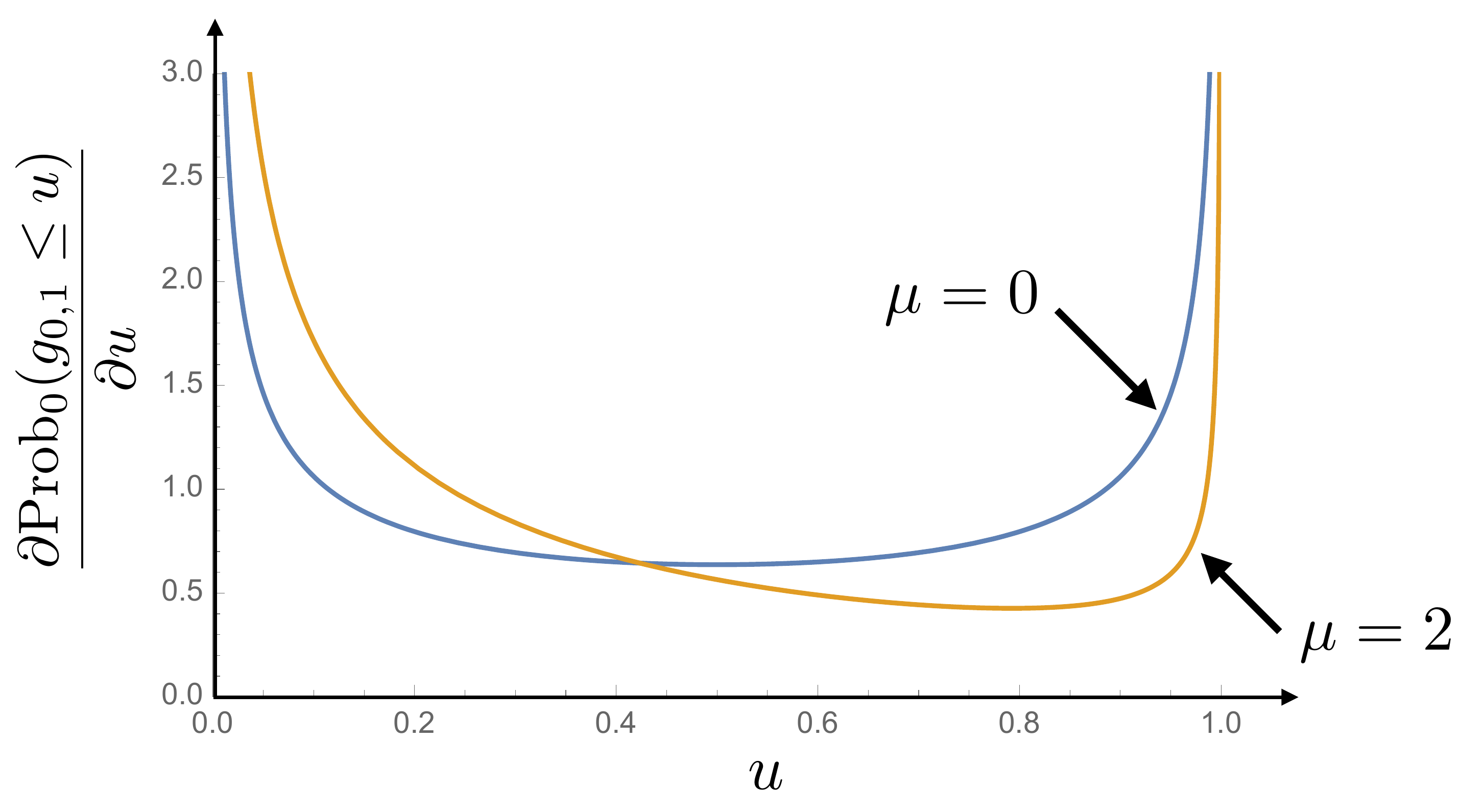}
\caption{Plot of the probability density of the last-passage time $g_{0,1}$ at $x_\star=0$, starting from $x_0=0$, on the unit time interval $[0,1]$ 
for the Brownian motion with a constant drift $\mu$, as given in Eq. \eqref{formula_BM_drift}, for two different values of the drift $\mu=2$ and 
$\mu=0$ [in the latter case case, the density is given by the arcsine law \eqref{ProbBrownian00}]. In both cases we set $D=1$.}\label{Fig_NM_mu}
\end{figure}
The fundamental solutions of interest are
\be \label{Phi_BM_drift}
\phi_\lambdaR(x)= \ed^{-x[ \mu + \sqrt{\mu^2 + 4 \lambda D}]/(2D)}
 \quad\textrm{and}\quad
\phi_\lambdaL(x)= \ed^{x[ -\mu + \sqrt{\mu^2 + 4 \lambda D}]/(2D)}.
\ee
Therefore, by injecting these expressions \eqref{Phi_BM_drift} in Eqs. \eqref{mRmLdefinition} and \eqref{mdefinition}, we get 
\be\label{m_lambda_0_BM_drift}
m(\lambda;0) = \mR(\lambda ; 0) - \mL(\lambda ; 0)=-\frac{1}{D} \sqrt{\mu^2 + 4 \lambda D}.
\ee
By substituting this formula \eqref{m_lambda_0_BM_drift} in Eq. (\ref{DoubleLTexpression}) and using standard results for 
inverse Laplace transforms (see e.g. \cite{Bateman1954}), we obtain
\be
\frac{\partial \Prob_{0}(g_{0,t} \leq u)}{\partial u}=
 \frac{1}{\pi}\frac{\ed^{-\frac{\mu^2 t}{4D} }}{\sqrt{u (t-u)}} 
+ \frac{\vert \mu\vert}{\sqrt{4 D}} \frac{\ed^{-(\mu^2/4D) u}}{\sqrt{\pi u}} 
\erf\left( \frac{\vert \mu\vert}{\sqrt{4D}} \sqrt{t-u} \right) \;, \label{formula_BM_drift}
\ee
where the the error function  is given in \eqref{erfdef}. A plot of this function is shown in Fig. \ref{Fig_NM_mu}. In particular, its asymptotic behaviors are given by
\bea\label{asympt_BM_mu}
\frac{\partial \Prob_{0}(g_{0,t} \leq u)}{\partial u} =
\begin{cases}
&\dfrac{1}{\sqrt{u}}\left[  \dfrac{\ed^{-\frac{\mu ^2 t}{4 D}}}{\pi 
   \sqrt{t}} +  \dfrac{\left| \mu \right|  \text{Erf}\left(\frac{\sqrt{t} \left| \mu \right| }{2
   \sqrt{D}}\right)}{\sqrt{4\pi D} }  
   \right] + O(\sqrt{u}) \;,\; u \to 0 \\
& \\
& \dfrac{1}{\sqrt{t-u}} \dfrac{\ed^{-\frac{\mu^2 t}{4D}}}{\pi \sqrt{t}} + O(\sqrt{t-u})\;, \; \hspace*{2.5cm} u \to t \;.    
   \end{cases}
\eea
Note that since $|\mu| >0$, the expression in (\ref{formula_BM_drift}) has a finite limit as $t\to+\infty$ (using ${\rm erf}(x \to \infty) = 1$). This means that $g_{0,t}$ converges at large time to a limiting random variable $g_{0, \infty}$ such that 
\be
\label{ProbBMmuinfty}
\frac{\partial \Prob_{0}(g_{0,\infty} \leq u)}{\partial u}=\frac{\vert \mu\vert}{\sqrt{4 D}} \frac{\ed^{-(\mu^2/4D) u}}{\sqrt{\pi u}}  \;.
\ee
On a physical basis this result is expected. Indeed, since the process is transient, at a certain random time $g_{0,\infty}$, it will never return to the origin. 

In the case $\mu=0$, which corresponds to the standard Brownian motion (which is of course not transient but recurrent), this formula \eqref{formula_BM_drift} reduces to 
\be
\label{ProbBrownian00}
\frac{\partial \Prob_{0}(g_{0,t} \leq u)}{\partial u}=
 \frac{1}{\pi\sqrt{u (t-u)}} \;,
\ee
which is the famous arc-sine law which also gives the occupation time density of the positive real half-axis \cite{Levy,Feller}.
It also coincides with the formula obtained by setting $x_\star=x_0$ in the
the expression~\eqref{lastBrown}. 

\subsection{Hyperbolic tangent force}
\label{TanhProcess}

The case $F(x)=2nD \tanh x$ where $n\in{\mathbb{N}}$ leads to reflectionless potentials \cite{CC} and is thus  explicitly solvable.  In the case $n=1$, first studied by Hongler \cite{Hongler1979}, one gets
\be
\phi_\lambdaR(x)=\frac{1}{\cosh x} \ed^{-x \sqrt{1+\lambda/D}} \;,
\ee
and $\phi_\lambdaL(x)=\phi_\lambdaR(-x)$. Then
\be
\mR(\lambda ; 0)=-\sqrt{1+\frac{\lambda}{D}}.
\ee
By analogy with the calculations in the case of the constant drift we get
\be
\label{Probtanh}
\frac{\partial \Prob_{0}(g_{0,t} \leq u)}{\partial u}=
 \frac{1}{\pi}\frac{\ed^{- D t}}{\sqrt{u (t-u)}} 
+ \sqrt{\frac{D}{\pi}} \frac{\ed^{-D u}}{\sqrt{u}} 
\erf\left( \sqrt{D(t-u)} \right).
\ee
Note that this is the same law as in the case of Brownian motion with drift $\mu=2 D$. 

As for the Brownian with a non-zero drift, the expression in  Eq.~\eqref{Probtanh} has a finite limit. Again, this means that $g_{0,t}$ converges at large time to a limiting random variable $g_{0, \infty}$ such that 
\be
\label{Probtanhinfty}
\frac{\partial \Prob_{0}(g_{0,\infty} \leq u)}{\partial u}=
  \sqrt{\frac{D}{ \pi u}} \ed^{-D u}.
\ee
On a physical basis this result is expected. Indeed, since the process is transient, at a certain random time $g_{0,\infty}$, it will never return to the origin.

The case $F(x)=-2D\tanh x$ can also be studied but the corresponding process is now recurrent. One obtains

\be
\mR(\lambda ; 0)=-\frac{\lambda/D}{\sqrt{1+\lambda/D}},
\ee
which gives 
\be
\label{ProbtanhOpposite}
\frac{\partial \Prob_{0}(g_{0,t} \leq u)}{\partial u}=
 \frac{1}{\pi}\frac{\ed^{- D t}}{\sqrt{u (t-u)}} 
+ \sqrt{\frac{D}{\pi}}  \frac{\ed^{-D (t-u)}}{\sqrt{ t-u }} 
\erf\left( \sqrt{D u} \right).
\ee
Note that the probability density differs from \eqref{Probtanh} by the exchange $u \leftrightarrow t-u$, an observation which we will come back to  in subsection \ref{sec:DualityProperties}.

\subsection{Ornstein-Uhlenbeck process}
\label{OUprocess}

\begin{figure}
\centering
\includegraphics[width = \linewidth]{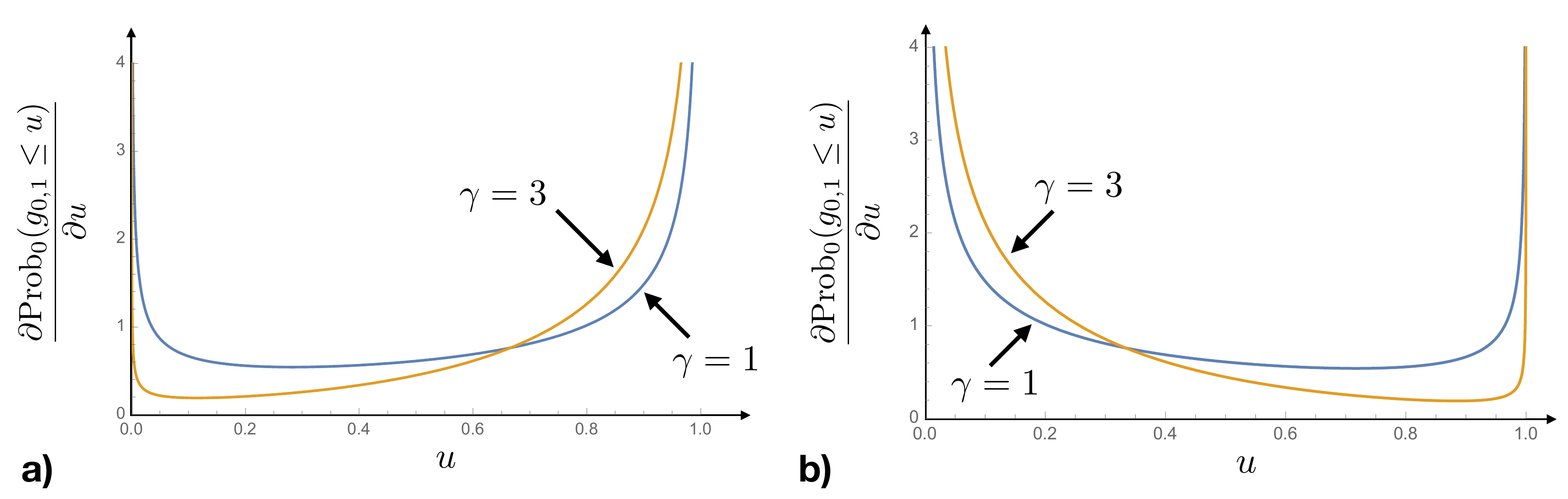}
\caption{{\bf a)}: Plot of the probability density of the last-passage time $g_{0,1}$ at $x_\star=0$, starting from $x_0=0$, on the unit time interval $[0,1]$ 
for the Ornstein-Uhlenbeck process, i.e. $F(x) = - \gamma x$, as given in Eq. \eqref{ProbOU}, for two different values of $\gamma=1$ and 
$\gamma=3$. {\bf b)}: Plot of the probability density of the last-passage time $g_{0,1}$ at $x_\star=0$, starting from $x_0=0$, on the unit time interval $[0,1]$ 
for the inverted Ornstein-Uhlenbeck process, i.e. $F(x) = \gamma x$, as given in Eq. \eqref{RepProbOU}, for two different values of $\gamma=1$ and 
$\gamma=3$. In both panels we set $D=1$. A comparison of the two panels clearly shows the symmetry $u\leftrightarrow t-u$ between the two problems.}\label{Fig_OU}
\end{figure}

For the Ornstein-Uhlenbeck process $F(x)=- \gamma x$ with $\gamma>0$
 and the generator \eqref{Generator} reads
\be
\Gen_x=D \frac{ \partial^2}{\partial x^2} -\gamma x \frac{\partial}{\partial x}.
\ee
A straightforward derivation would start from the Dirichlet kernel, constructed by the image method, and then use \eqref{relProbTxstarGDxstar}. One can as well use the general formalism.
The fundamental solution of interest is
\be
\phi_\lambdaR(x)= \ed^{(\gamma/4 D)x^2} U\left(\frac{\lambda}{\gamma} -\frac{1}{2}, x\sqrt{\frac{\gamma}{D} }\right)
\ee
where $U(a,x)$  is the  parabolic cylinder function (see for instance \cite{INST}), and  $\phi_\lambdaL(x)=\phi_\lambdaR(-x)$. 
One finds
\be
\frac{\mR(\lambda ; 0)}{\mR(\lambda + \lambda')}=2^{\frac{\lambda'}{\gamma}}
\frac{\Gamma\left(\frac{\lambda}{\gamma}\right)}{\Gamma^2\left(\frac{\lambda}{2\gamma}\right)}
\frac{\Gamma^2\left(\frac{\lambda+\lambda'}{2\gamma}\right)}{\Gamma\left(\frac{\lambda+\lambda'}{\gamma}\right)}.
\ee
Finding the inverse Laplace transform is still feasible and gives
\be
\label{ProbOU}
\frac{\partial \Prob_{0}(g_{0,t} \leq u)}{\partial u}=
 \frac{2  \gamma}{\pi} \frac{ \ed^{- \gamma (t-u)}}
{ \sqrt{\left[1-\ed^{-2  \gamma u} \right]\left[1-\ed^{-2  \gamma(t-u)} \right]}}.
\ee
A plot of this probability density is shown in Fig. \ref{Fig_OU} a) for two different values of $\gamma$. In particular, we see that the probability that
``$g_{0,t}$ is close to $t$'' increases with $\gamma$, which is expected since the trajectory of the particle gets more and more confined
close to $x=0$ as $\gamma$ is increased. This observation can be made more quantitative by analysing the asymptotic behaviors of the function in \eqref{ProbOU} which read
\bea \label{asympt_OU}
\frac{\partial \Prob_{0}(g_{0,t} \leq u)}{\partial u}=
\begin{cases}
&\dfrac{1}{\sqrt{u}} \dfrac{\ed^{-\gamma t/2}}{\pi} \sqrt{\dfrac{\gamma}{\sinh(\gamma t)}} + O (\sqrt{u}) \;, \hspace*{1cm} \; u \to 0 \;,\\
& \\
&\dfrac{1}{\sqrt{t-u}} \dfrac{\ed^{\gamma t/2}}{\pi} \sqrt{\dfrac{\gamma}{\sinh(\gamma t)}} + O(\sqrt{t-u}) \;, \;\;\; u \to t  \;.
\end{cases}
\eea
In fact the result in \eqref{ProbOU} can be readily derived by making a suitable change of time 
\be
t\rightarrow  T(t)=\frac{ \ed^{2\gamma t}-1}{2 \gamma} \;,
\ee
which maps the Ornstein-Uhlenbeck process $\dd x/\dd t=-\gamma x + \sqrt{2D} \, \eta(t)$ to the standard Brownian problem
$\dd X /\dd T = \sqrt{2D}\, \eta(T)$. One can check that 
\be
x(t)=\ed^{-\mu t} X\left( \frac{ \ed^{2\gamma t}-1}{2 \gamma} \right)
\ee
and by using \eqref{ProbBrownian00} the result \eqref{ProbOU} follows immediately.

It is also instructive to study the inverted Ornstein-Uhlenbeck, i.e. the repulsive case $F(x)=  \gamma x$ with $\gamma>0$. In this case, we obtain
\be
\label{RepProbOU}
\frac{\partial \Prob_{0}(g_{0,t} \leq u)}{\partial u}=
 \frac{2  \gamma  }{\pi} \frac{\ed^{-   \gamma u}}
{ \sqrt{\left[1-\ed^{-2  \gamma  u} \right]\left[1-\ed^{-2 \gamma  (t-u)} \right]}}.
\ee
As in the previous examples \eqref{ProbBMmuinfty} and \eqref{Probtanhinfty}, we notice that in the limit $t\to\infty$, the density goes to a finite limit which means that $g_{0,t}\rightarrow g_{0,\infty}$. Its density is given by
\bea \label{mOU_inf}
\frac{\partial \Prob_{0}(g_{0,\infty} \leq u)}{\partial u} =  \frac{2  \gamma  }{\pi} \frac{\ed^{-   \gamma u}}
{ \sqrt{\left[1-\ed^{-2  \gamma  u} \right]}} \;.
\eea	
A plot of this probability density is shown in Fig. \ref{Fig_OU} b) for two different values of $\gamma$. Note also that, as before in \eqref{Probtanh} and \eqref{ProbtanhOpposite}, the probability densities \eqref{ProbOU} and \eqref{RepProbOU} differ by the exchange $u\leftrightarrow t-u$ (see also Fig. \ref{Fig_OU}), an observation which we shall come back to in subsection \ref{sec:DualityProperties}. Consequently, the asymptotic behaviors of the probability density in \eqref{RepProbOU} can be read off from \eqref{asympt_OU} by changing $u$ in $t-u$. 

\subsection{Bessel process reflected at the origin}

We consider a Bessel process with the diffusion constant  $D=1/2$.
\be \label{Bessel_Gen}
\Gen_x=\frac{1}{2} \frac{ \partial^2}{\partial x^2} +\frac{1-2\mu}{2x} \frac{ \partial}{\partial x}
\ee
with $0<\mu<1$ and reflected at the origin. One obtains
\be
\phi_\lambdaR(x)= \left( x \sqrt{2\lambda}\right)^\mu K_\mu(x \sqrt{2\lambda})
\ee
\be
\frac{\phi'_\lambdaR(x)}{\phi_\lambdaR(x)} \underset{x\to 0}{\sim} -\frac{\Gamma(1-\mu)}{\Gamma(\mu)}
\left( \frac{x}{2}\right)^{2\mu-1}(2\lambda)^\mu.
\ee
Therefore
\be
\frac{m(\lambda ; 0)}{m(\lambda+\lambda')}=\left(\frac{\lambda}{\lambda+\lambda'} \right)^\mu.
\ee
By inverse Laplace transform one gets
\be\label{Prob_Bessel}
\frac{\partial \Prob_{0}(g_{0,t} \leq u)}{\partial u}=\frac{\sin(\pi \mu)}{\pi} u^{\mu-1} (t-u)^{-\mu}.
\ee
For a general discussion of this result,  see \cite{Bertoin} and references therein. In particular, for $\mu=1/2$, we recover the arcsine law, which is somehow expected from Eq. \eqref{Bessel_Gen} (note however that there is a reflecting boundary condition at $x=0$). However, it is interesting to note that, for $\mu \neq 1/2$, the asymptotic behavior of this probability density (\ref{Prob_Bessel}) as $u \to 0$ (respectively as $u \to t$) diverges with an exponent $\mu$ (respectively $1-\mu$) which is different from $1/2$, and thus different from the cases studied above [see Eqs. \eqref{asympt_BM_mu} or \eqref{asympt_OU}], and in particular different  from the arcsine law \eqref{ProbBrownian00}.

\section{Case of a force deriving from an even potential with $x_0=x_\star=0$}
In this section we continue our study of the case where the initial and target point coincide. It is shown that a certain duality relation between diffusions is the key to explain the symmetry properties satisfied by last-passage densities under the exchange $u\leftrightarrow t-u$ observed above. Furthermore we show that the basic objects which enter into  \eqref{DoubleLTexpression} have an interesting spectral interpretation in the context of the Weyl theory. As an interesting application we show that the small-$t$ expansion of the expected last-passage time involves the Korteveg-de Vries invariants.

\subsection{Duality properties for a symmetric potential}
\label{sec:DualityProperties}

It is convenient to express the density of last-passage time in the following form
\be
\label{DensityF}
\frac{\partial \Prob_{0}(g_{0,t} \leq u)}{\partial u}= h(t-u)f(u).
\ee
The decomposition of the density in terms of two functions $h(t)$ and $f(t)$ is made explicit on the various examples which have been treated above. For instance in the case of the hyperbolic tangent force \eqref{Probtanh} one has
\be
f(t)=\frac{\ed^{-Dt}}{\sqrt {\pi t}} 
\ee
\be
h(t)=\frac{\ed^{-Dt}}{\sqrt {\pi t}}+ \sqrt{D} \, \erf\left( \sqrt{D t} \right).
\ee
As seen on \eqref{ProbtanhOpposite}, under the change $F(x)\rightarrow \tilde{F}(x)=-F(x)$, and denoting  the corresponding last-passage time in $x_\star = x_0$ by $\tilde g_{0,t}$,  the functions $f(t)$ and $h(t)$ are exchanged, namely
\be
\label{DensityOppositeF}
\frac{\partial \Prob_{0}(\widetilde{g}_{0,t} \leq u)}{\partial u}= \widetilde{h}(t-u)\tilde{f}(u)
\ee
where $\tilde{h}=f $ and $ \widetilde{f}=h$. This property which we observed in a few particular cases above turns out to be generic for any arbitrary even potential $U(x) = U(-x)$ (see Appendix \ref{Duality} for details). 

In fact the time-reversal symmetry which connects \eqref{DensityF} and \eqref{DensityOppositeF} arises from the following duality between processes driven by opposite forces.
Let  $x(\tau)$ and $\widetilde{x}(\tau)$ be two independent diffusion processes starting from the origin and evolving via the overdamped Langevin equations
\be
\label{Langevin2Eq}
\dd x(\tau)= F(x)\, \dd \tau + \sqrt{2D}\,\dd B(\tau),
\ee
\be
\label{Langevin3Eq}
\dd \widetilde{x}(\tau)= -F(\widetilde{x})\, \dd \tau + \sqrt{2D}\,\dd \widetilde{B}(\tau)
\ee
where $B(\tau)$ and $\widetilde{B}(\tau)$ are two independent Brownian motions.
Let $g_{0,t}$ and $\widetilde{g}_{0,t}$ be  the corresponding last-passage times at the origin before $t$. When the potential is even, i.e. $U(x)=U(-x)$, one can prove 
(see appendix C) that
 $\widetilde{g}_{0,t}\overset {\textrm{law}}=t-g_{0,t}$.  Stated differently if $h(t-u)f(u)$ is the density of $g_{0,t}$, the density of $\widetilde{g}_{0,t}$ will be $h(u)f(t-u)$. The diffusion processes  $x(\tau)$ and $\widetilde{x}(\tau)$ are said to be dual (or conjugate). This concept of duality  has been used in several contexts, in particular non-intersecting diffusions \cite{Assio} and continued fractions \cite{Comtour}, just to mention a~few.
 
\subsection{A trace formula for the mean last-passage time}

In this subsection  we explore the relationship between the mean last-passage time and the so-called Weyl coefficient which is  a central object in the Sturm-Liouville theory. Using this correspondence, we derive a trace formula, which expresses the mean last-passage time as a difference of two partition functions. This result is illustrated on some  examples.
For simplicity we shall restrict ourselves to the case of an even potential, although our approach has a much wider scope. A thorough presentation making extensive use of the trace formula derived in \cite{Gesztesy} will be discussed in a separate publication.

In the case of an even potential $U(x)=U(-x)$, the right and left eigenfunctions are related by $\phi_\lambdaR(x)=\phi_\lambdaL(-x)$, and according to the definitions of the various logarithmic derivatives in Eqs. \eqref{mRmLdefinition} and \eqref{mdefinition}, one has, in this case, $m(\lambda ; x_\star)=2\mL(\lambda ; x_\star)$. Then, for $x_0=x_\star=0$,  the expression \eqref{MeanValue} for the Laplace transform of the mean value of the last-passage time becomes 
\be
\label{LTmeanValueUeven}
\int_0^{+\infty} \ed^{-\lambda t} \, \mathbb{E}_{0}\left[g_{0, t} \right]\, \dd t=
\frac{1}{\lambda}
\frac{\dd  \ln \mR(\lambda ; 0) }{\dd \lambda} \;,
\ee
where $\mR(\lambda ; x)$ is defined in \eqref{mRmLdefinition} in terms of the non-decreasing eigenfunction of the diffusion generator $\phi_\lambdaR(x)$   which obeys the boundary condition on the right.

\subsubsection{Relation between the mean last-passage time and the Weyl coefficient}

In Appendix \ref{KernelExpression} we recall how to express the transition kernel for the Langevin equation \eqref{LangevinEq} in terms of the eigenfunctions of the Hamiltonian $H$ given in \eqref{Hdefinition}, namely
\be 
\label{H_text}
H=-D\frac{\dd^2}{\dd x^2}+V(x)
\ee
with
\be
\label{relVandU}
V(x)=\frac{1}{4D} \left[ \frac{\dd U}{\dd x} \right]^2 -\frac{1}{2}\frac{\dd^2 U}{\dd x^2}.
\ee
The  fundamental solutions $\phi_\lambdaR$ and $\phi_\lambdaL$ of the diffusion generator are related to the eigenfunctions $\psi_ {E,R}$ and $\psi_ {E,L}$ of the Hamiltonian with $E=-\lambda \in\mathbb{R^-}$ according to \eqref{relphiRpsiR} and \eqref{relphiLpsiL}. If the potential is even and differentiable, then $U'(0)=0$, and therefore
\be
\label{mRUeven}
\mR(\lambda;0)=\left.\frac{\psi'_{-\lambdaR}(x)}{\psi_{-\lambdaR}(x)}\right\vert_{x=0}.
\ee
The logarithmic  derivative in \eqref{mRUeven} is in fact related to the so-called ``Weyl function'' $\WeylR(E)$, a central object of the Sturm-Liouville theory,  whose definition is briefly recalled  below.

For every $E\in\mathbb{C}$, any solution of the Schr\"odinger equation $H\psi=E\psi$ can be constructed as a linear combination of two independent solutions $\psi_{E,1}(x)$ and $\psi_{E,2}(x)$ which satisfy the Cauchy problem


\be
\label{SturmLiouvilleBC}
\begin{cases}
&\psi_{E,1}(0)=1 \\
&\psi'_{E,1}(0)=0
\end{cases}
\qquad\textrm{and}\qquad
\begin{cases}
&\psi_{E,2}(0)=0 \\
&\psi'_{E,2}(0)=1
\end{cases} .
\ee 
Under suitable conditions (see for instance \cite{Coddington} or \cite{Messiah1961}), 
 there exists a unique linear combination
\be
\label{decompsiWeyl}
\psi_ {E,R}(x)= \psi_{E,1}(x) + \WeylR(E) \, \psi_{E,2}(x) 
\ee
which is in $L^2(\mathbb{R^{+}})$.
The corresponding solution is called the Titchmarch-Weyl function and the logarithmic derivative
\be
\label{defWeyl}
 \WeylR(E)= \left.\frac{\psi'_{E,R}(x)}{\psi_{E,R}(x)}\right\vert_{x=0} 
\ee
is the Weyl coefficient. For the Schr\"odinger equation on the full line there exist two Weyl functions  $\psi_R$ and $\psi_L $ which  are respectively in $L^2(\mathbb R^+)$ and $L^2(\mathbb R^-)$.  In the case of the half-line problem, or if  the potential  is even, it is sufficient to work with $ \WeylR(E)$.
Since $\mR(\lambda;0)$ is related to the logarithmic derivative of   $\psi_\lambdaR$ at $x=0$ by \eqref{mRUeven}, the right Weyl function coincides with  $\mR(\lambda;0)$ for $\lambda \in \mathbb{R^+}$
\be
 \WeylR(-\lambda)=\mR(\lambda;0) \;.
 \ee
 Recall that  $\mR(\lambda ; x)$ is defined in \eqref{mRmLdefinition} in terms of the non-increasing eigenfunction of the diffusion generator $\phi_\lambdaR(x)$   which obeys the boundary conditions on the right.
 As a consequence  the Laplace transform of the mean value of the last-passage time  becomes 
\be
\label{LTmeanValueUevenBis}
\int_0^{+\infty} \ed^{-\lambda t} \, \mathbb{E}_{0}\left[g_{0, t} \right]\, \dd t=
\frac{1}{\lambda}
\frac{\dd  \ln  \WeylR(-\lambda) }{\dd \lambda},
\ee
and subsequently the Laplace transform of its rate of variation with respect to the observation time $t$ reads
\be
\label{LTrate}
\int_0^{+\infty} \ed^{-\lambda t} \, \frac{\dd}{\dd t }\mathbb{E}_{0}\left[g_{0, t} \right]\, \dd t=
\frac{\dd  \ln  \WeylR(-\lambda) }{\dd \lambda}.
\ee
 
 \subsubsection{A trace formula}

When the potential  $V(x)$  is even, the eigenstates of $H$ defined on the whole real axis,  can be classified by their parity and the spectral problem is reduced to a 
Schr\"odinger problem on the positive  half-axis  with Dirichlet or Neumann boundary conditions at $x=0$ . The Green's functions of the corresponding Hamiltonians $H_\textrm{D}$  and  $H_\textrm{N}$, defined for $E\in\mathbb{C \setminus R^+}$,  are denoted by  $\widehat{g}_\textrm{D} (x,y ; E)$ and $\widehat{g}_\textrm{N}  (x,y ;E)$ respectively. They can be expressed in terms of the previous eigenfunctions of $H$ (see for instance \cite{Craig}). For our purpose we need only to consider the case  $x\ge y>0$ where
 \be
 \label{DirichletGreenf}
 \widehat{g}_\textrm{D} (x,y ; E)= \frac{1}{D} \psi_{E,2}(y) \psi_{E,R}(x) 
 \ee
 and
  \be
 \label{NeumannGreenf}
 \widehat{g}_\textrm{N} (x,y; E)=-\frac{1}{D \WeylR(E)} \psi_{E,1}(y) \psi_{E,R}(x).
 \ee
 The difference between the Neumann  and Dirichlet Green's functions at coinciding points reads
 \be
 \label{diffgNgD}
 \widehat{g}_\textrm{N}(x,x; E)- \widehat{g}_\textrm{D} (x,x; E)=-\frac{1}{ D\WeylR(E)}
\left[ \psi_{E,R}(x)\right]^2 .
 \ee
Since $\psi_ {E,R}(x)$ is an eigenfunction of $H$ given in \eqref{H_text} with eigenvalue $E$,
 \be
\left[ \psi_{E,R}(x)\right]^2=-D\frac{ \partial}{ \partial x} \left(
\psi_{E,R}\frac{\partial^2 \psi_{E,R}}{\partial x \,\partial E}-\frac{\partial \psi_{E,R}}{\partial x} \frac{\partial \psi_{E,R}}{\partial E} 
\right)
= -D\frac{ \partial}{ \partial x} \left( 
\left[ \psi_{E,R}(x)\right]^2 \frac{\partial^2 \ln\psi_{E,R}(x)}{\partial x \, \partial E}
 \right) \;.
 \ee
As a consequence, since $\psi_{E,R}(x)$ is integrable when $x\to +\infty$, the difference between the integrated Green's functions at coinciding points reads
\be
\int_0^{+\infty} \left[ \widehat{g}_\textrm{N} (x,x; E)-  \widehat{g}_\textrm{D} (x,x; E)\right] \, \dd x= -\frac{1}{ \WeylR(E)} 
\left.\left[ \psi_{E,R}(0)\right]^2 \frac{\partial^2 \ln\psi_{E,R}(x)}{\partial x \, \partial E}\right\vert_{x=0} \;.
\ee
Moreover, $\psi_{E,R}(x)$ obeys the Sturm-Liouville boundary conditions given in  \eqref{SturmLiouvilleBC} and eventually, according to the definition \eqref{defWeyl} of the Weyl coefficient, we get
\be
\label{intdiff}
 \int_0^{+\infty} \left[ \widehat{g}_\textrm{N} (x,x; E)-  \widehat{g}_\textrm{D} (x,x; E)\right] \, \dd x=  
 - \frac{\partial \ln \WeylR(E)}{\partial E}.
\ee

The relation \eqref{intdiff} can be rewritten in terms of operator traces by noticing that for $E=-\lambda \in \mathbb{R^-}$
\be
 \int_0^{+\infty} \left[ \widehat{g}_\textrm{N} (x,x; E)-  \widehat{g}_\textrm{D} (x,x; E)\right] \, \dd x= 
 \int_0^{+\infty} \Tr \left( \ed^{-t H_\textrm{N}} - \ed^{-t H_\textrm{D}}  \right)  \ed^{-\lambda t}\dd t
\ee
By comparison with the Laplace transform in \eqref{LTrate}  one gets that the variation rate of the mean value for the last-passage time takes the form
\be
 \label{RateMeanValue}
 \frac{\dd}{\dd t }\mathbb{E}_{0}\left[g_{0, t} \right]=  \Tr \left( \ed^{-t H_\textrm{N}} - \ed^{-t H_\textrm{D}}  \right)  \;.
\ee

If the spectrum of $H$ is discrete, then $\Tr\, \ed^{-t H_\textrm{N}}=\Tr_{+} \, \ed^{-t H} $, where  $\Tr_{+}$ denotes a partial  trace over the even eigenstates. Similarly  $\Tr \, \ed^{-t H_\textrm{D}}$ coincides with the partial trace over the odd eigenstates, namely  $\Tr \,\ed^{-t H_\textrm{D}}=\Tr_{-} \, \ed^{-t H} $, where  $\Tr_{-}$ denotes a trace over the odd eigenstates. Therefore \eqref{RateMeanValue} becomes
  \be
 \label{RateMeanValueBBis}
 \frac{\dd}{\dd t }\mathbb{E}_{0}\left[g_{0, t} \right]=  {\Tr}_{+} \, \ed^{-t H}  -\Tr_{-} \,\ed^{-t H} \;.
\ee

 The traces in  \eqref{RateMeanValue} or \eqref{RateMeanValueBBis} play the role of  `` quantum partition functions''.
In order to obtain the mean value of the last-passage time one can first calculate the latter partition functions and then get the mean value of interest by integrating the relation \eqref{RateMeanValueBBis} with the initial condition $\lim_{t\to 0} \mathbb{E}_{0}\left[g_{0, t} \right]=0$. 

  \be
\label{gmean00}
 \mathbb{E}_{0}\left[g_{0, t} \right]
 =t + \sum_{n\neq 0}\left[\frac{1-\ed^{-E_n^+ t}}{E_n^+} - \frac{1-\ed^{-E_n^- t}}{E_n^-}\right],
\ee
 The term linear in $t$ comes from the ground state $E_{0}=0$, and $E_n^+$ ($E_n^-$) denotes the $n$th even (odd) energy. Another derivation of this result,  using an identity given in \cite{Voros1,Voros2}, is presented in Appendix D.

\subsubsection{Examples and remarks}
 
Let us first  return to the case of the Ornstein-Uhlenbeck process  discussed in subsection \ref{OUprocess}. In this case, according to \eqref{H_text}-\eqref{relVandU}, the corresponding Hamiltonian is that of the harmonic oscillator shifted by a constant
 \be \label{H_OU}
 H= -D \frac{\partial^2}{\partial x^2} +\frac{\gamma^2}{4D}x^2 -\frac{\gamma}{2} \;.
 \ee
 The eigenvalues of   $H$  read $E_n=\gamma(n+\frac{1}{2})-\frac{\gamma}{2}=\gamma n$ with $n\in \mathbb{N}$. Therefore
 \be
 \Tr_{+}\, \ed^{-t H}  =\frac{1}{1-\ed^{-2\gamma t}} 
 \ee
 and
 \be
 \Tr_{-} \,\ed^{-t H} =\frac{\ed^{-\gamma t}}{1-\ed^{-2\gamma t}}  \;.
 \ee
Since the mean value vanishes at the initial time, Eq.~\eqref{RateMeanValueBBis} gives
 \be \label{exact_OU}
 \mathbb{E}_{0}\left[g_{0, t} \right] = \frac{1}{\gamma} \ln\left(\frac{1 + \ed^{\gamma t}}{2} \right) \;.
 \ee
Note that obtaining this result \eqref{exact_OU} from the probability density given in Eq. \eqref{ProbOU} is actually not so simple.

An alternative expression for \eqref{RateMeanValue} involving Schr\"odinger operators on the whole real axis  is
\be
 \label{RateMeanValue4}
 \frac{\dd}{\dd t }\mathbb{E}_{0}\left[g_{0, t} \right]=  \Tr \left( \ed^{-t H} -\ed^{-t H_{D,0}} \right) \;.
\ee
 In this expression $H$ is the Schr\"odinger  operator on $L^2(\mathbb R)$ and $H_{D,0}$ the Schr\"odinger operator with Dirichlet boundary condition at the origin defined on $L^2(\mathbb R^{-})\oplus L^2(\mathbb R^{+})$. The right hand side of \eqref{RateMeanValue4} has been studied in depth in \cite{Simon,Rybkin} where it is shown that the coefficients of the small-time expansion are related to the Korteveg-de Vries invariants. Using these results one obtains the small-$t$ expansion of the last-passage time
\be
 \label{RateMeanValue3}
 \mathbb{E}_{x}\left[g_{x, t} \right]=  \frac{t}{2}-\frac{t^2}{4} V(x)-\frac{t^3}{12} \left({D}\,\frac{V''(x)}{2}-V^2 (x)\right)+O(t^4) \;.
\ee 
This formula is valid for any starting point $x$ and any smooth diffusion process on the line. In particular, specifying this formula \eqref{RateMeanValue3} to the Ornstein-Uhlenbeck process corresponding to $V(x) = \gamma^2x^2/(4D) - \gamma/2$ [see Eq. \eqref{H_OU}] and setting $x=0$, one obtains $ \mathbb{E}_{x}\left[g_{x, t} \right] = t/2 + \gamma t^2/2 + O(t^4)$, which coincides with the direct small-$t$ expansion of the exact formula in \eqref{exact_OU}.

\section{Limit of an infinite observation time for a transient process}
\label{InfiniteObservationTime}

The previous examples  in subsections \ref{TanhProcess} and \ref{OUprocess} invite us to treat within our framework the general case of a transient process with exit at $+\infty$ and/or $-\infty$. Our aim in this section is to determine the last-passage time  limiting density when the observation time  $t$   goes to infinity. 


When the observation time  $t$   goes to infinity the last-passage time  density for a transient process  becomes proportional to the transition kernel by virtue of its factorized expression \eqref{Product},
\be
\label{ProductBis}
\frac{\partial  \Prob_{x_0}(g_{x_\star,\infty} \leq u)}{\partial u}\propto 
 P (x_\star, u \vert x_0, 0)  \;.
\ee
In order to determine the prefactor we  notice that
 \be
 \lim_{t\to\infty} h_t(x_\star)
 =  \lim_{\lambda \to 0} \lambda \int_0^{+\infty} \ed^{-\lambda t} h_t(x_\star) \, \dd t \;.
\ee
and then, by virtue of  \eqref{Product}, the limit \eqref{ProductBis} reads
\be
\label{LimInfiniteDuration}
\frac{\partial  \Prob_{x_0}(g_{x_\star,\infty} \leq u)}{\partial u}=
 \left( \lim_{\lambda \to 0} \lambda \,\widehat{h}_{\lambda}(x_\star)\right) \,  P (x_\star, u \vert x_0, 0) 
\ee
where, according to \eqref{LTh},
\be
\label{LimitlambdahTozero}
\lim_{\lambda \to 0} \lambda \,\widehat{h}_{\lambda}(x_\star)=-
D\left[ \frac{ \phi'_\zeroR(x_\star)}{\phi_\zeroR(x_\star)}-  \frac{\phi'_\zeroL(x_\star)}{\phi_\zeroL(x_\star)}\right].
\ee
By using the definition of $m(\lambda,x_\star)$ in \eqref{mdefinition}, the last-passage time density can therefore be rewritten for any transient process as
\be
\label{LimInfiniteDurationBis}
\frac{\partial  \Prob_{x_0}(g_{x_\star,\infty} \leq u)}{\partial u}= D\, m(0;x_\star)
 P (x_\star, u \vert x_0, 0) .
\ee

In fact the expression of the multiplicative coefficient in \eqref{LimInfiniteDurationBis}  can be expressed explicitly in terms of the potential $U(x)$. 
In order to do this we must determine the zero modes of the diffusion generator $\Gen$ with the proper boundary conditions. The function  $\phi_\zeroL$  is non-decreasing and obeys the left  boundary  condition, while $\phi_\zeroR$ is non-increasing and obeys the right boundary condition. According to \eqref{Generateur} 
they both satisfy the equation
\be
\left[D \frac{\partial^2}{\partial y^2} -\frac{\dd U}{\dd y} \frac{\partial }{\partial y}\right]\phi_0(y)=0 \;.
\ee
We now consider the following four cases:

\begin{itemize}

\item[$\bullet$] 1) For a diffusion along the real axis which is transient with exit  at $+\infty$, $\int_{y}^{+\infty} \ed^{U(x)/D}\, \dd x<+\infty$ and $\int_{-\infty}^{y} \ed^{U(x)/D}\, \dd x = +\infty$, there exists a decreasing solution  which vanishes when $y\to+\infty$, namely $\phi_\zeroR(y)=\int_y^{+\infty}\ed^{U(x)/D}\, \dd x$, while the solution which is non-increasing on the left  is  a  constant, namely $\phi_\zeroL(y)=Cste$.  Therefore, according to \eqref{LimInfiniteDuration} and \eqref{LimitlambdahTozero},
\be
\label{LimInfiniteDurationCase1}
\frac{\partial  \Prob_{x_0}(g_{x_\star,\infty} \leq u)}{\partial u}=
D \frac{\ed^{U(x_\star)/D}}{\int_{x_\star}^{+\infty} \ed^{U(x)/D}\, \dd x}
\,  P (x_\star, u \vert x_0, 0) 
\ee

\item[$\bullet$]  
2) For a  diffusion along the real axis  which is  transient with exit  at $-\infty$,  $\int_{-\infty}^{y} \ed^{U(x)/D}\, \dd x<+\infty$ and $\int_y^{+\infty}\ed^{U(x)/D}\, \dd x=+\infty$, there exists an increasing solution $\phi_\zeroL(y)$ which vanishes when $y\to-\infty$, namely $\phi_\zeroL(y)=\int_{-\infty}^y\ed^{U(x)/D}\, \dd x$, while the  solution which is non-increasing on the right is $ \phi_\zeroR(y)=Cste$.  In this case, according to \eqref{LimInfiniteDuration} and \eqref{LimitlambdahTozero}, one has
\be \label{transient_case2}
\frac{\partial  \Prob_{x_0}(g_{x_\star,\infty} \leq u)}{\partial u}=
D \frac{\ed^{U(x_\star)/D}}{\int_{-\infty}^{x_{\star}} \ed^{U(x)/D}\, \dd x} 
\,  P (x_\star, u \vert x_0, 0) \;.
\ee

\item[$\bullet$]
3) If the diffusion is transient on both sides, which happens if $\int_{-\infty}^{+\infty} \ed^{\frac{U(x)}{D}} \dd x<+\infty$, one gets
\be \label{transient_case3}
\frac{\partial  \Prob_{x_0}(g_{x_\star,\infty} \leq u)}{\partial u}=
D\left[ \frac{1}{\int_{-\infty}^{x_\star} \ed^{U(x)/D}\, \dd x}
+\frac{1}{\int_{x_\star}^{+\infty} \ed^{U(x)/D}\, \dd x}\right] \ed^{U(x_\star)/D}
\,  P (x_\star, u \vert x_0, 0)  \;.
\ee

\item[$\bullet$]
4) For a diffusion along the positive real half-axis  which is transient with exit  at $+\infty$ and a reflection at $y=0$ the non-decreasing solution $\phi_\zeroL(y)$ whose derivative vanishes at $y=0$ is a constant  function : $\phi'_\zeroL(y)=0$ for all $y$ and then
\be
\label{LimInfiniteDuration4}
\frac{\partial  \Prob_{x_0}(g_{x_\star,\infty} \leq u)}{\partial u}=
D \frac{\ed^{U(x_\star)/D}}{\int_{x_\star}^{+\infty} \ed^{U(x)/D}\, \dd x}
\,  P (x_\star, u \vert x_0, 0) \;.
\ee

\end{itemize}
One can check that our formulae are in agreement with  the general case treated  in  \cite{Pitman,Salminen1997}.

Finally, we notice that the Laplace transform of the last-passage time density in the infinite observation time limit can be determined from the relation
\be
\mathbb{E}_{x_0}\left[\ed^{-\lambda' g_{x_\star, \infty} }\right]=  \lim_{\lambda \to 0} \lambda \int_0^{+\infty} \ed^{-\lambda t} 
 \, \mathbb{E}_{x_0}\left[\ed^{-\lambda' g_{x_\star, t}} \right]\, \dd t
\ee
and the expression \eqref{DoubleLTBis}. We get
\be
\label{LTdensityinfinitetime}
\mathbb{E}_{x_0}\left[\ed^{-\lambda' g_{x_\star, \infty} }\right]=
\frac{m(0 ; x_\star)}{m(\lambda' ; x_\star)}\, \mathbb{E}_{x_0}\left[\ed^{-\lambda'T_{x_\star} }\right]
\ee

The mean last-passage time  can be derived from the Laplace transform of the last-passage time density given in \eqref{LTdensityinfinitetime}
 with the result
\be
\label{meangmeanTtinfinite}
 \mathbb{E}_{x_0}\left[g_{x_\star, \infty} \right]=
\left.\frac{\dd  \ln m(\lambda ; x_\star) }{\dd \lambda}\right\vert_{\lambda=0}+\mathbb{E}_{x_0}\left[T_{x_\star}\right] \blue\;, \;\;\; {\rm for} \;\;\; x_0 < x_\star \;.
\ee

\section{Reflected Brownian motion with a constant drift}\label{sec:drift}

\subsection{Semi-classical interpretation of the transition kernel}
\label{secsemiclassical}

In this section we derive an explicit expression of the transition kernel for the  Brownian motion with drift reflected at the origin. We are not just rederiving a well known result, see for instance  \cite{BORO}, instead our approach provides an interpretation in terms of classical paths which will be used for the multi-particle problem.
The starting point is to express the transition kernel $P(x,t \vert x_0, 0)$ in terms of a Schr\"odinger Green's function.
According to \eqref{relToSchroedinger}-\eqref{Hdefinition}, when $U(x)=-\mu x$
\be
\label{transi}
P(x,t \vert y, 0)=\ed^{\frac{\mu (x-y)}{2D}}\langle y \vert \ed^{-tH} \vert x\rangle
\ee
where the Hamiltonian reads
\be
H= -D \frac{ \partial^2}{\partial x^2} +\frac{\mu^2}{4D}.
\ee
The Green's function $\langle y \vert \ed^{-tH} \vert x\rangle$  is symmetric under the exchange of $x$ and $y$.
Following Refs. \cite{LINET} and \cite{Montex}, it can be computed by using a complete set of eigenstates of $H$.
Since the problem is free, we can take a set of plane waves 
\be
\psi_k(x)=   \left[ \ed^{-ikx}+r(k) \ed^{ikx} \right]
\ee
parametrized by their momentum $k$ and energy $E_{k}=Dk^2+\mu^2/(4D)$. The reflection coefficient $r(k)$ is fixed by the boundary conditions. These are inherited from the diffusion process which obeys reflecting boundary conditions at the origin : the probability current vanishes at the origin so that, by virtue of~\eqref{DefJP},
\be
D\frac{\partial P}{\partial x}-\mu P=0.
\ee
Therefore one gets
\be
r(k)= \frac{-2ikD-\mu}{-2ikD+\mu},
\ee
and the Schr\"odinger Green's function takes the form
\be
\langle y \vert \ed^{-tH} \vert x\rangle
=\int_{0}^{\infty} \psi_{k} (y) \psi^*_{k}(x)\ed^{-E_{k}t} \,\dd k \;.
\ee
After some rearrangements it reads 
\be
\label{Green}
\langle y \vert \ed^{-tH} \vert x\rangle= \frac{1}{\sqrt {4\pi Dt}}\ed^{-\frac{(x-y)^2}{4Dt}-\frac{\mu^2 t}{4D}}+\frac{1}{\sqrt {4\pi Dt}}\ed^{-\frac{(x+y)^2}{4Dt}-\frac{\mu^2 t}{4D}}+\frac{1}{2\pi}\int_{{-\infty}}^{\infty} [r(k)-1] \,\ed^{ik(x+y)-E_{k}t} \, \dd k \;.
\ee
Using a set of identities discovered by Gautschi \cite{Gautschi}, one can show that for $z>0$
\be
\label{Gautshi}
\frac{i}{\pi}\int_{-\infty}^{+\infty}\frac{{\rm e}^{-u^2}}{iz+u} \,{\rm d} t= {\rm e}^{z^2} {\erfc} (z) \;.
\ee
By using \eqref{Gautshi} with $z = (x+y+\mu t)/\sqrt{4 Dt}$ , the last term of \eqref{Green} can be expressed in terms of the complementary error function, and the transition kernel \eqref{transi} takes the final form
\bea
\label{propa}
P(x,t \vert y, 0)&=& \frac{\ed^{\,\mu \frac{x-y}{2D}}}{\sqrt {4\pi Dt}} \left[ \ed^{-\frac{(x-y)^2}{4Dt}-\frac{\mu^2 t}{4D}}+
\ed^{-\frac{(x+y)^2}{4Dt}-\frac{\mu^2 t}{4D}}\right]
 \,\, -\frac{\mu}{2D}\ed^{\frac{\mu x}{D}}\erfc\left(\frac{x+y+\mu t}{\sqrt{4Dt}}\right).
\eea

In several cases the Green 's function and the transition kernel can be interpreted in terms of classical paths. However in the presence of singularities such as wedges or corners such expansions have to be modified. This  reflects the fact that there exist so-called diffractive orbits which obey the law of classical mechanics everywhere except on the singularities of the potential (see for instance \cite{Bog}). In our simple one-dimensional setting, there are  only two classical paths connecting $y$ and $x$ in a given time $t$, namely the direct path and the one which is reflected at the origin.
However the  transition kernel \eqref{propa} is not simply given as a sum of two contributions. There are indeed three terms which have the following interpretation:
\begin{itemize}
\item[$\bullet$] The first exponential term involves the action 
$
S_1=(x-y-\mu t)^2 /(4Dt)
$
 of the direct path travelled at a constant velocity $v=(x-y)/t$. 
\item[$\bullet$] Similarly the  second exponential term involves the action  
$
S_2=(x+y+\mu t)^2/(4Dt)- \mu x/D
$
 of the reflected path travelled at a constant velocity  $v=(x+y)/t$.

\item[$\bullet$] The last term should be interpreted as a diffractive contribution which arises because there is an infinite potential barrier at $ x=0$ and the particle is scattered non-classically. The reflection coefficient which enters in \eqref{Green}  gives a finite contribution when $\mu\neq 0$.
Another interpretation, of a more probabilistic nature, is suggested in \cite{Clark} in which the last term reflects the fact that the particle spends a certain finite time, the so-called local time, at the origin.
\end{itemize}

\subsection{Mean value of the first-passage time}
\label{secmeanTdriftrefl}

The mean value of the first  passage time can be obtained from 
\be
\mathbb{E}_{x_0}\left[T_{x_\star}\right]=- \left.\frac{\partial}{\partial \lambda} \mathbb{E}_{x_0}\left[\ed^{-\lambda T_{x_\star}}\right]\right\vert_{\lambda=0} \;,
\ee
by using  \eqref{EspTx0supxstar} and \eqref{EspTx0infxstar}, which requires the determination of the fundamental solutions $\phi_\lambdaR(x)$ and $\phi_\lambdaL(x)$ of the diffusion generation $\Gen_x$  given in \eqref{Gendrift}. They are  linear combinations of $\ed^{-\rho_+ x}$ and $\ed^{-\rho_- x}$, where $\rho_\pm$ is the solution of the quadratic equation $D \rho^2 - \mu \rho - \lambda = 0$ 
namely
\be
\label{defrhopm}
-\rho_\pm= \frac{1}{2D}\left[\mp \sqrt{\mu^2+ 4D\lambda}-\mu\right] \;.
\ee
Therefore $\phi_\lambdaR(x)$, the decreasing function which vanishes when $x$ goes to infinity, reads (up to a multiplicative constant)
\be
\label{philRdriftrefl}
 \phi_\lambdaR(x)=  \ed^{-  \rho_+ x} \;,
 \ee
while  $\phi_\lambdaL(x)$, the increasing function whose derivative vanishes at $x=0$ can be chosen as
 \be
 \label{philLdriftrefl}
 \phi_\lambdaL(x)= \rho_+\ed^{- \rho_-x}-\rho_-\ed^{- \rho_+x} \;.
\ee
We get 
\bea
\mathbb{E}_{x_0}\left[T_{x_\star}\right]
&=&
\frac{x_\star-x_0}{\mu} + \frac{D}{\mu^2}\left[ \ed^{-(\mu/D)x_\star}-\ed^{-(\mu/D)x_0}\right] \;,
 \quad\textrm{if $x_0<x_\star$}\quad
 \\ 
\label{meanTinfinitx0supxstar}
\mathbb{E}_{x_0}\left[T_{x_\star}\right] &=&\frac{x_0-x_\star}{\mu} \ed^{-(\mu/D)(x_0-x_\star)} \;,
 \quad\hspace*{2.7cm}\textrm{if $x_0>x_\star$}\quad \;,
\eea
thus recovering standard results.

\subsection{Last-passage time density in an infinite observation time limit}
\label{secLastpassagedensityinfinitegeneric}

\begin{figure}
\centering
\includegraphics[width = 0.7\linewidth]{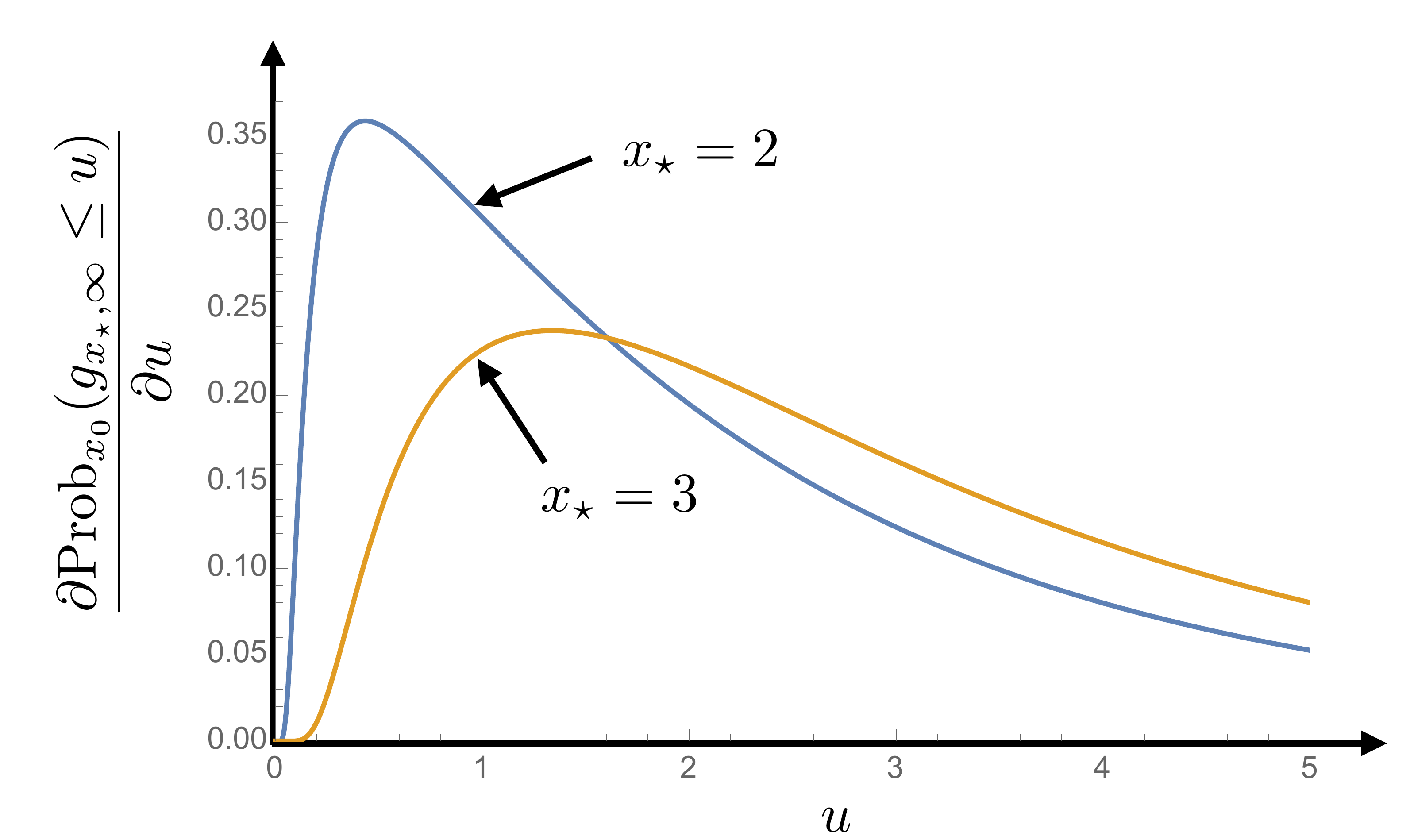}
\caption{Plot of the probability density of the last-passage time $g_{x_\star,\infty}$, for the reflected Brownian motion, starting at $x_0=1$ with constant drift $\mu = 1$ and diffusion constant $D=1$, in the case of an infinite observation time, as given in Eq.~\eqref{densityprobagdriftrefl}. The two curves correspond to two different values of the target point $x_\star = 2$ and $x_\star = 3$.}\label{Fig_mu_reflected}
\end{figure}

Setting  $U(x)=-\mu x$ in \eqref{LimInfiniteDuration4} we get
 \bea
 \label{densityprobagdriftrefl}
&&\frac{\partial  \Prob_{x_0}(g_{x_\star,\infty} \leq u)}{\partial u}=
\mu
\,  P (x_\star, u \vert x_0, 0) \;, \nonumber \\
&&= \mu\left(\frac{\ed^{\,\mu \frac{x_\star-x_0}{2D}}}{\sqrt {4\pi Du}} \left[ \ed^{-\frac{(x_\star-x_0)^2}{4Du}-\frac{\mu^2 u}{4D}}+
\ed^{-\frac{(x_\star+x_0)^2}{4Du}-\frac{\mu^2 u}{4D}}\right]
 \,\, -\frac{\mu}{2D}\ed^{\frac{\mu x_\star}{D}}\erfc\left(\frac{x_\star+x_0+\mu u}{\sqrt{4Du}}\right) \right) \;,
\eea
where we have used the expression of $P (x_\star, u \vert x_0, 0)$ given in \eqref{propa}. 
A plot of this probability density is shown in Fig. \ref{Fig_mu_reflected}. From the explicit expression in \eqref{densityprobagdriftrefl}, we obtain the asymptotic behavior, for $x_0<x_*$
\bea \label{asympt_BM_ref}
\frac{\partial  \Prob_{x_0}(g_{x_\star,\infty} \leq u)}{\partial u} \approx
\begin{cases}
& \mu \dfrac{\ed^{\mu\frac{x_\star-x_0}{2D}}}{\sqrt{4 \pi D u}}\, \ed^{-\frac{(x_\star-x_0)^2}{4Du}} \;, \; u \to 0 \\
& \\
& \frac{2\sqrt{D}}{\mu\sqrt{\pi u^3}} \ed^{-\frac{\mu^2u}{4D}+\frac{\mu(x_\star-x_0)}{2D}}(1+\frac{\mu(x_\star+x_0)}{2D}+\frac{\mu^2x_\star x_0}{4D^2})\;, \; \hspace*{0.6cm} u \to \infty \;.
\end{cases}
\eea

For $x_0<x_\star$, 
the mean value of the last-passage time can be conveniently obtained from the relation \eqref{meangmeanTtinfinite} with the result
\be
\label{meangmeanTmeandif}
\mathbb{E}_{x_0}\left[g_{x_\star,\infty}\right]=\frac{2D}{\mu^2}-\frac{D}{\mu^2}\ed^ {-\frac{\mu x_\star}{D}}+ \mathbb{E}_{x_0}\left[T_{x_\star}\right].
\ee

According to \eqref{meanTinfinitx0supxstar} the mean last-passage time thus reads
\be
\label{meanginfinitedriftrefl}
\mathbb{E}_{x_0}\left[g_{x_\star, \infty} \right]=\frac{x_\star-x_0 }{\mu} + \frac{2D}{\mu^2}-  \frac{D}{\mu^2} \ed^{-\frac{\mu x_0}{D}}
 \quad\textrm{if $x_0<x_\star$}\;.
\ee
This formula will be used in the next section.



\section{Application to the emptying time of a box}

In this section, we show how the results on last-passage times obtained in the previous sections, and
in particular the ones concerning the reflected Brownian motion with a constant drift in Section \ref{sec:drift},
can be used to study the statistics of the emptying time of a box.

\subsection{Emptying time for a generic diffusion process}

\begin{figure}
\centering
\includegraphics[width = 0.7 \linewidth]{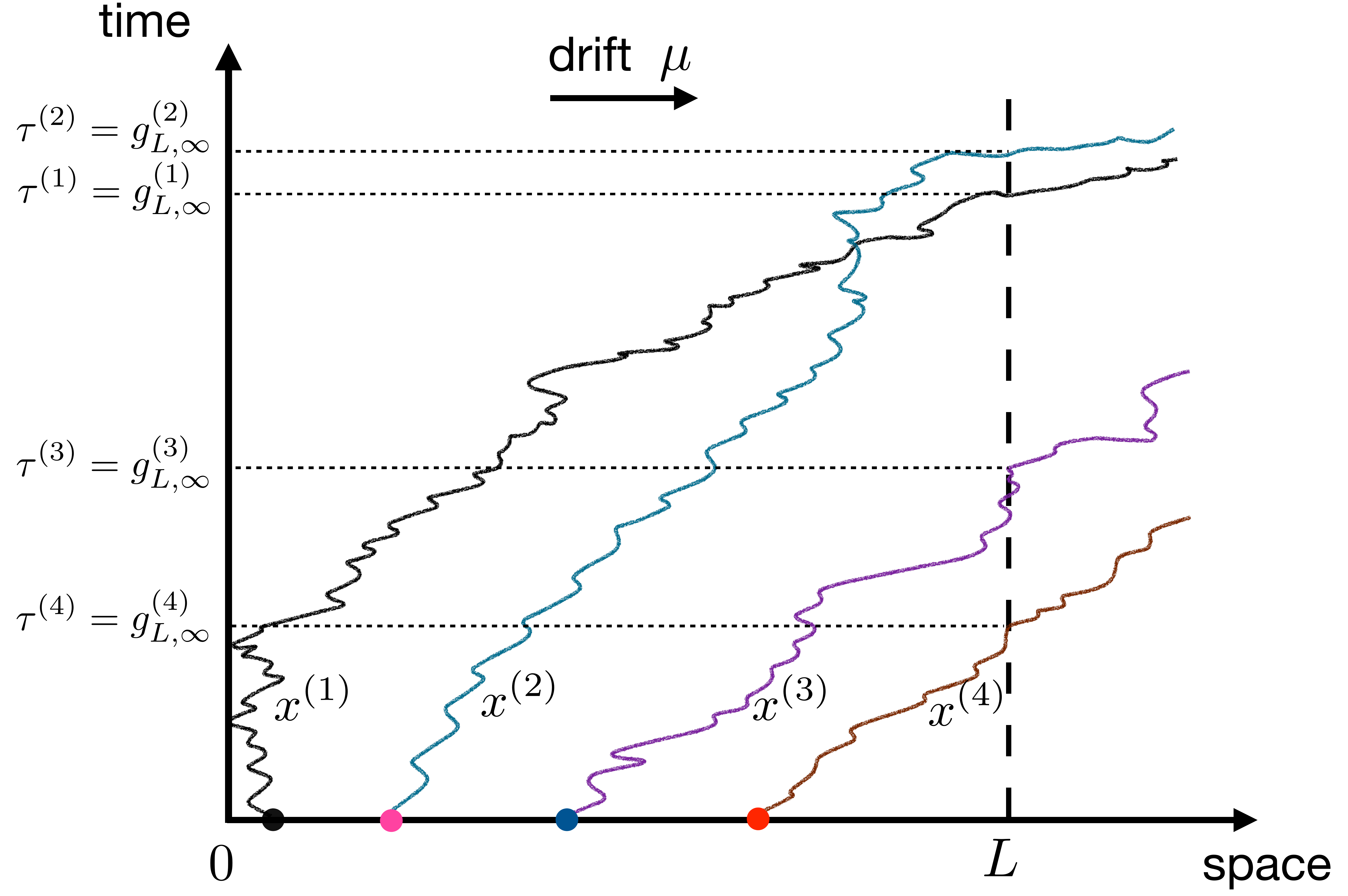}
\caption{Trajectories of $N=4$ non-interacting Brownian particles $x^{(1)}(t), \cdots, x^{(N)}(t)$ in the presence of a positive drift $\mu$ with a reflecting boundary condition at $x=0$. Initially, the particles are uniformly distributed inside the box $[0,L]$ and we denote by $\tau^{(i)} = g^{(i)}_{L,+\infty}$ the last-passage time at $x=L$ of the $i$-th particle. The emptying time of the box $T_{\rm vac}$ is given by $T_{\rm vac} = \max\{\tau^{(1)}, \cdots, \tau^{(N)}\} = \max\{g^{(1)}_{L,\infty}, \cdots, g_{L,\infty}^{(N)} \}$ [see Eq. \eqref{def_Tvac_max}], here $T_{\rm vac} = \tau^{(2)}=g^{(2)}_{L,\infty}$.}\label{Fig_vacuum}
\end{figure}

We consider a set of $N$ independent and non-interacting particles $x^{(1)}(t), \cdots, x^{(N)}(t)$, whose initial positions $x^{(i)}(t=0) = x^{(i)}_0$ are distributed inside the region $[0,L]$  with  probability density $\prod_{i=1}^N P(x_0^{(i)};t=0)$ where the probability law is independent of the particle label. In the following, we will mainly consider the case where the particles are initially uniformly distributed over $[0,L]$, which corresponds to $P(x_0^{(i)};t=0) = 1/L$, for all $i=1, \cdots, N$. At later times $t>0$, the particles move inside the region $[0,+\infty[$. If they are subjected to an external potential $U$ such that $\int_{x}^{+\infty} \ed^{U(x)/D}\, \dd x<+\infty$ for any $x\geq 0$, then the process is  transient  with exit at $+\infty$ and the particles will leave the region $[0,L]$ forever with probability one. {When the observation time is infinite} the emptying time of the region $0<x <L$, denoted by  $\Tvac$, is  the first time when all particles are at positions $x^{(i)}(t=T_{\rm vac})>L$ and will never cross the abscissa $L$ again. In other words $\Tvac$ is the maximum of the last-passage times for the $N$ particles. As a consequence, if  the last-passage time for the particle with label $i$ in a realization of the process is denote by $g^{(i)}_{L,\infty}$, the cumulative distribution function of $\Tvac$ is given by
\be\label{def_Tvac_max}
\Prob\left( \Tvac \leq \tau\right)=
\Prob\left(g^{(i)}_{L,\infty} \leq \tau \;, \;  \forall i\in\{1\ldots,N\}\right) \;.
\ee
For a given realization of the initial positions, the study of the statistics of $\Tvac$ thus amounts to compute the distribution of the maximum of a set
of $N$ independent but {\it non-identically distributed} random variables. Interestingly, extreme value questions concerning independent but non-identically distributed arose recently in various problems, in particular in random matrix theory \cite{LGMS2018} or in the quantum mechanics of trapped fermions~\cite{DDMS2017}.

{Here the initial positions $x_0^{(i)}$ of the particles are randomly distributed on $[0,L]$ and, to make progress, one would like to 
average the distribution of $\Tvac$ over the initial conditions. In analogy with disordered systems where a realization of the disorder plays an analogous role to the initial condition in our problem, it was argued by Derrida and Gerschenfeld \cite{DG09}, in the related albeit different context of current fluctuations in diffusive systems (see also \cite{BMRS20} for a similar study in the context of active particles) that one has to distinguish between two different ways of 
averaging over the initial conditions: (i) the annealed average, where the probability distribution of $\Tvac$ is averaged over {\it all} the realizations of the initial condition 
and (ii) the quenched average where the probability distribution is computed for the {\it typical} realizations of the initial configurations. The annealed and quenched averages are defined respectively~as 
\begin{eqnarray}
&&P_{\rm an}(\tau) \, =  \overline{ \Prob\left( \Tvac \leq \tau\right) } \;, \label{def_ann} \\
&&P_{\rm qu}(\tau) = \exp{\left[\overline{\ln({\Prob\left( \Tvac \leq \tau\right)})}\right]} \;, \label{def_quen}
\end{eqnarray}
where $\overline{\cdots}$ denotes an average over the initial conditions. It turns out that, for our problem, in the scaling limit, considered here, of a large number of particles, $N \to \infty$, and large box size $L \to \infty$ at fixed density $\rho = N/L$, the typical behaviors of $P_{\rm an}(\tau)$ and $P_{\rm qu}(\tau)$ are actually similar (they differ only at the level of large deviations) and therefore, in the following we will only present the computation for the annealed average $P_{\rm an}(\tau)$, and relegate the analysis of $P_{\rm qu}(\tau)$ to Appendix~\ref{app_quenched}}.

Starting from Eq. (\ref{def_ann}), the annealed average $P_{\rm an}(\tau)$ can be written in a quite simple way, by taking advantage of the fact that 
the particles are independent and identically distributed in the initial state, i.e.,
\be
\label{CumulativeTout}
P_{\rm an}(\tau) =  \overline{\Prob\left(\Tvac \leq \tau\right)}= \left[\overline{p}(\tau) \right]^N
\ee
where $\overline{p}(\tau)$ is an average over the initial position of a particle, 
\be
\label{defp}
\overline{p}(\tau)= \int_0^{L}\Prob_{x}(g_{L,\infty} \leq \tau) P(x;t=0) \, \dd x \;.
\ee
For a transient process, a path starting from $x \in [0,L]$ crosses $L$ {in a finite time} with probability one during an infinite observation time, i.e., {$\Prob_{x}(T_L <+\infty)=1$. Therefore the coefficient of the Dirac mass in \eqref{decomPi} vanishes and
 $\int_0^\infty \partial \Prob_{x}(g_{L,\infty}\leq u)/\partial u \, \dd u = 1$}. Moreover since the single particle probability density at the initial time  is normalized, i.e., $\int_0^LP(x;t=0)\, \dd x =1$, we can check from \eqref{defp} that
\be
\lim_{\tau \to +\infty} \overline{p}(\tau)=1 \;.
\ee
To analyse the large-$\tau$ limit of $P_{\rm an}(\tau)$, it is convenient to introduce
$
\overline{q}(\tau)=1- \overline{p}(\tau)
$
and write the cumulative distribution function of the emptying time as
\be\label{expr_Tvac_N}
P_{\rm an}(\tau) = \overline{\Prob\left( \Tvac \leq \tau\right)}= \left[1- \overline{q}(\tau)\right]^N \;.
\ee
Hence, after averaging over the initial condition, one is back to the problem of extreme statistics of $N$ independent and identically distributed random
variables with an effective cumulative distribution $\overline{p}(\tau) = 1 - \overline{q}(\tau)$. Nevertheless, the problem remains non-trivial because here $\overline{q}(\tau)$ depends on an additional parameter, namely $L$, which itself depends also on $N$ since we will be interested in the limit where $N/L = \rho$ is fixed.

For the sake of conciseness, we introduce the notation $\Pi_{x,L}(u)$ for the probability density function (PDF) of $g_{L,\infty}$, i.e., 
\be
\Pi_{x,L}(u)=\frac{\partial  \Prob_{x}(g_{L,\infty} \leq u)}{\partial u}
\ee
in terms of which we have
\be
\label{defqbar}
\overline{q}(\tau)= \int_0^L q(\tau,x)  P(x;t=0) \, \dd x \;,\;\; {\rm where} \;\; q(\tau,x) =\int_\tau^{+\infty} \Pi_{x,L}(u)\, \dd u \;.
\ee
During an infinite observation time the PDF $\Pi_{x,L}(u)$ is proportional to the transition kernel and, since the process in question corresponds to case 4) of Section \ref {InfiniteObservationTime}, we get from \eqref{LimInfiniteDuration4} 
\be
\label{qtauxinfinite}
q(\tau,x)=
 \frac{D}{\int_{L}^{+\infty} \ed^{[U(y)-U(L)]/D}\, \dd y} \,
\int_\tau^{+\infty} P (L, u \vert x, 0) \, \dd u \;,
 \ee
 where $P (L, u \vert x, 0)$ is the transition kernel, which can be expressed in terms of the eigenfunctions of the generator (see Appendix \ref{KernelExpression} for details). This expression \eqref{qtauxinfinite}, together with \eqref{expr_Tvac_N} and \eqref{defqbar} allows to compute, in principle, the cumulative distribution of $T_{\rm vac}$ for $N$ particles, for any arbitrary transient process with exit at $+\infty$ and any initial distribution of the positions.

\subsection{Emptying time for a single reflected Brownian motion with a constant drift}

In the following, we will focus on the case of the reflected Brownian motion with a constant drift, i.e. $U(x)=-\mu x$ {with $\mu>0$}, and we first consider the case of a single particle, which can be studied in detail thanks to the results presented in Section \ref{sec:drift}. In particular, in view of future applications to $N$ particles, we derive the asymptotic form of the PDF $\Pi_{x,L}(u)$ in the limit of large $L$.

Specializing the general formula (\ref{qtauxinfinite}) to the case $U(x)=-\mu x$ we obtain
\be
q(\tau,x)=\mu \int_\tau^{+\infty} P (L, u \vert x, 0) \, \dd u \;,
\ee
where the transition kernel $P (L, u \vert x, 0)$ has been computed previously and is given in \eqref{propa}. The PDF $\Pi_{x,L}(u) = \mu P (L, u \vert x, 0)$ can thus be naturally split into three contributions 
\be
\label{PDF_explicit}
\Pi_{x,L}(u)=\mu\left[P_1 (L, u \vert x, 0)+P_2 (L, u \vert x, 0)+P_3 (L, u \vert x, 0)\right]
\ee
with
\be
\label{defP1}
P_1 (L, u \vert x, 0)= \frac{1}{\sqrt{4\pi D u}} \ed^{-(L-x-\mu u)^2/(4Du)} \;,
\ee
\be
\label{defP2}
P_2 (L, u \vert x, 0)= \frac{1}{\sqrt{4\pi D u}} \ed^{\mu L/D} \ed^{-(L+x+\mu u)^2/(4Du)} \;,
\ee
\be
\label{defP3}
P_3 (L, u \vert x, 0)=-\frac{\mu}{2D}\ed^{\mu L/D}\erfc\left(\frac{L+x+\mu u}{\sqrt{4 D u}}\right)\ \;,
\ee
where $P_1$ is the constant drift kernel (associated to direct paths from $x$ to $L$), $P_2$ is its image kernel with respect to $x=0$ (associated to paths which have been reflected at $x=0$) and $P_3$ is a diffractive contribution as discussed in the context of the semi-classical interpretation 
in subsection \ref{secsemiclassical}. 

In the following, we consider the case where the particles are initially distributed uniformly over $[0,L]$, i.e. $P(x_0^{(i)};t=0) = 1/L$ for all $i=1, \cdots, N$ and, in the limit of large $L$, it is thus natural to study $\Pi_{x,L}(u)$, with $x = y L$, with $y = O(1)$ for large $L$ (and $0\leq y \leq 1$). To get an idea of the large $L$ limit of the PDF $\Pi_{x = y L,L}$ it is useful to analyse the mean last-passage time which can be read off from \eqref{meanginfinitedriftrefl} by setting $x_ 0 = x = y L$ and $x_\star = L$. It reads
\bea \label{mean_LPT_largeL}
\mathbb{E}_{x=y L}(g_{L,\infty}) &=& \frac{L}{\mu}(1-y) + \frac{2D}{\mu^2} - \frac{D}{\mu^2} \ed^{-\frac{\mu y L}{D}} \\
&=& \frac{L}{\mu}(1-y) + O(1) \;, \; {\rm as} \;\;\; L \to \infty \;.
\eea
Hence, for large $L$, the average last-passage time is of order ${O}(L)$ and is simply the time needed for a particle moving at a constant velocity $\mu$ to travel a distance $L-x = L(1-y)$, which corresponds to the distance to travel to escape from the box, starting from $x$. From the semi-classical interpretation of the transition probability, this term clearly comes from the direct paths, described by the kernel $P_1$ in \eqref{defP1}. 
\begin{figure}[ht]
\centering
\includegraphics[width = 0.7\linewidth]{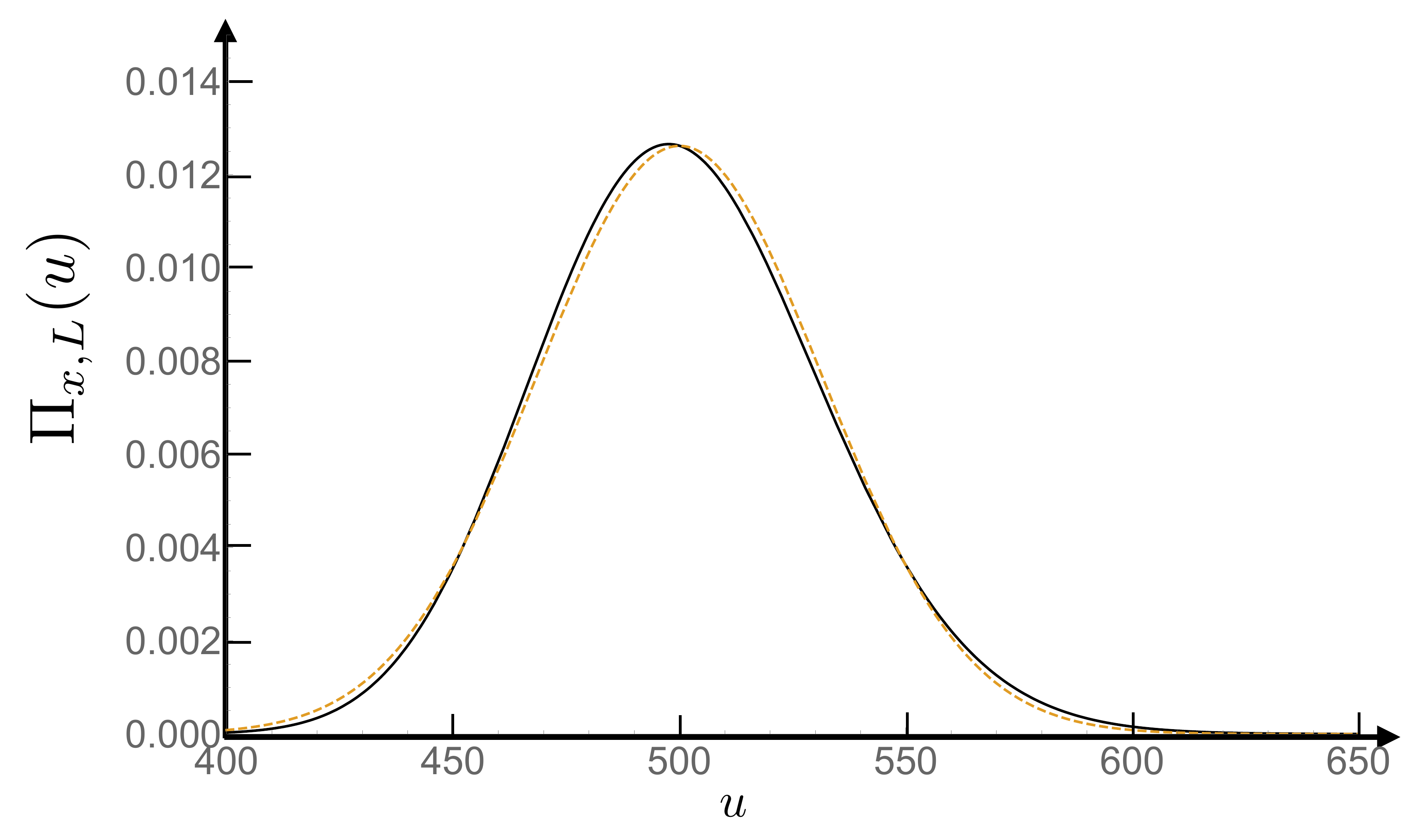}
\caption{Plot of $\Pi_{x,L}(u)$ as given in Eqs. \eqref{PDF_explicit}-\eqref{defP3} (solid black line) compared to the Gaussian scaling form (dotted orange line), valid for large $L$, given in Eq. \eqref{scaling_Pi} for $D=1$, $\mu=1$, $x=1$ and $L=500$.} \label{Fig_Pi}
\end{figure}

In fact, in the limit of large $L$, one can show that the leading contribution to the full PDF $\Pi_{x=yL,L}(u)$ is given by $P_1$. In this limit, one can then show that $\Pi_{x=yL}(u)$ takes the scaling form 
\begin{eqnarray}\label{scaling_Pi}
\Pi_{x,L}(u) \underset{L \to \infty}{\sim} \sqrt{\frac{\mu^3 }{{D\,L}}} F\left[\sqrt{\frac{\mu^3}{{D\,L}}}\left(u - \frac{1}{\mu}(L-x), \frac{x}{L}\right)\right] \;, \;\;\; F(z,y) = \frac{1}{\sqrt{4 \pi (1-y)}} \ed^{-\frac{z^2}{4(1-y)}} \;,
\end{eqnarray} 
which shows that the fluctuations of $g_{x, \infty}$ around the average ``deterministic'' or advective value ${\cal T}_{\rm ad}(x) = (L-x)/\mu$ are of order $O(\sqrt{L})$ and Gaussian. We see on this scaling form \eqref{scaling_Pi} that these fluctuations are controlled by a ``diffusive'' time scale ${\cal T}_{\rm diff} = \sqrt{DL/\mu^3}$. This time scale can be understood as follows. At a given large time $\tau$, the position of the Brownian motion in the presence of a drift $\mu$, starting from $x = L y$, can be estimated (neglecting the effects of the reflecting boundary) as $x(\tau) \approx L y + \mu \tau + \sqrt{2 D \tau} \chi_G$, where $\chi_G$ is a centred Gaussian random variable of unit variance. If one estimates $g_{x,\infty}$ by setting $x(g_{yL,\infty}) \approx L$, one finds, to leading order for large $L$
\bea\label{tau_diff}
g_{yL,\infty} - {\cal T}_{\rm ad} \approx {\cal T}_{\rm diff}\sqrt{2(1-y)} \tilde \chi_G \;, \;\;\; {\rm with}\;\;  {\cal T}_{\rm diff} = \sqrt{\frac{DL}{\mu^3}} \;,
\eea  
where $\tilde \chi_G = - \chi_G$ is also a centred Gaussian random variable of unit variance. Hence, this diffusive time scale appears naturally as a consequence of the fact that, to escape the box, the ``effective'' distance to be travelled, driven by the external force $\mu$, is not exactly $(L-x)$ but instead $(L-x) \pm \sqrt{2 D L/\mu}$, because of diffusion. In Fig.~\ref{Fig_Pi}, we compare the exact expression of $\Pi_{x,L}(u)$ to the Gaussian scaling form (\ref{scaling_Pi}) for $L=500$. This supports the fact that, for large $L$, the PDF $\Pi_{x,L}(u)$ is almost a Gaussian.  

One can then easily compute the variance from Eq. (\ref{scaling_Pi}), which is given by 
\begin{eqnarray}\label{asympt_variance}
{\rm Var}(g_{x=yL,\infty}) =  L \frac{2D(1-y)}{\mu^3} + o(L) = 2 {\cal T}_{\rm diff}^2 (1-y) + o(L)\;.
\end{eqnarray}
Note that a direct computation of the variance from the full expression in \eqref{PDF_explicit} shows that the corrections in \eqref{asympt_variance} are actually of order $O(1)$. In fact, one can show that all higher cumulants are also of order $O(L)$ for large $L$, namely
\be 
\mathbb{E}\left[ g_{x=y\,L,\infty}^p \right]^c = \frac{\mu L}{2D}(1-y) \left(\frac{2D}{\mu^2}\right)^p(2p-3)!!
+ o(L) \;,
\ee
where $n!! = n(n-2)(n-4)\cdots$ denotes the double factorial, with the convention $(-1)!!=1$. This yields back, for $p=1$, $\mathbb{E}\left[ g_{y\,L,L}\right]\approx L(1-y)/\mu$ as well as, for $p=2$, the result for the variance given in~\eqref{asympt_variance}. One further obtains $\mathbb{E}\left[ g_{x=y L,L}^3\right]_c  \approx 12 D^2 L(1-y)/\mu^5$ as well as $\mathbb{E}\left[ g_{x = y L,L}^4\right]_c\approx 120 D^3\,L(1-y)/\mu^7$. 

By inserting the scaling form for {$\Pi_{x,L}(u)$} given in Eq. \eqref{scaling_Pi} into Eq. \eqref{defqbar}, we obtain straightforwardly	 the scaling form of $q(\tau,x)$
\bea \label{scaling_qtaux}
q(\tau,x) \approx G\left[ \frac{1}{{\cal T}_{\rm diff}}\left(\tau - \frac{L-x}{\mu}\right), \frac{x}{L}\right] \;\;, \;\; G(z,y) = \frac{1}{2} {\rm erfc} \left( \frac{z}{2\sqrt{(1-y)}}\right) \;.
\eea
To proceed, we now need to average this expression over the initial condition to obtain $\bar{q}(\tau)$ given in Eq. \eqref{defqbar}. Here we focus on the case where the particles are initially uniformly distributed over $[0,L]$, i.e. we substitute $P(x;t=0) = 1/L$ in Eq. \eqref{defqbar}. We then see, from the first argument of the function $G$ in \eqref{scaling_qtaux} that the integral over $x$ in \eqref{defqbar} is actually dominated by the region of $x = O(\sqrt{L})$\footnote{This is indeed the case for a uniform initial density but this might not be the case for other choices of the initial density.} -- we remind indeed that ${\cal T}_{\rm diff} = O(\sqrt{L})$. Performing this integral, we find that $\bar{q}(\tau)$ can be written in the scaling form
\bea\label{scaling_qbar}
\bar{q}(\tau) \approx \frac{\mu {\cal T}_{\rm diff}}{L} H\left[ \frac{\tau - {\cal T}_{\rm ad}}{{\cal T}_{\rm diff}{}}\right] \;,
\eea
where ${\cal T}_{\rm diff}$ is given in \eqref{tau_diff} and where we used the notation ${\cal T}_{\rm ad} = {\cal T}_{\rm ad}(x=0) = L/\mu$. In Eq. \eqref{scaling_qbar}, the scaling function $H(z)$ is given by
\bea\label{expr_H}
H(z) = \int_z^\infty G(z',0) \, \dd z' = \sqrt{\frac{1}{\pi}}\, \ed^{-\frac{z^2}{4}} - \frac{z}{2} \, {\rm erfc} \left( \frac{z}{2}\right) \;.
\eea
Its asymptotic behaviors, which will be useful in the following, are given by
\be\label{asympt_Han}
H(z) \sim \begin{cases}
&\dfrac{2}{\sqrt{\pi} z^2} \, \ed^{-\frac{z^2}{4}} \;, \; \hspace*{1.cm} z \to +\infty \;,\\
& \\
& |z| \;, \; \hspace*{2.5cm} z \to -\infty \;.
\end{cases}
\ee
We will now use this scaling form \eqref{scaling_qbar} to study the emptying time for $N$ reflected Brownian motions with a drift. 

\subsection{Emptying time for $N$ reflected Brownian motions with a constant drift}\label{sec:Gumbel_an}

The starting point of our analysis in the case of $N$ Brownian motions with a drift is the exact relation given in Eq. \eqref{expr_Tvac_N}, which can be analysed, for large $N$, along the lines of what is usually done in extreme value statistics (see for instance \cite{Gum58} or \cite{MPS2020} for a more recent review). Here, we consider the limit where both $N \to \infty$ and $L \to \infty$ but keeping the ratio $N/L = \rho$ fixed. In this limit, $\bar{q}(\tau)$ is small [see Eq. \eqref{scaling_qbar}] and one can thus write, to leading order for large $N$ and $L$
\bea \label{re-exponentiate}
P_{\rm an}(\tau) = (1 - \bar{q}(\tau))^N \simeq \ed^{- N \bar{q}(\tau)} \;.
\eea 
By setting $N = \rho L$ and injecting the scaling form \eqref{scaling_qbar} in \eqref{re-exponentiate}, we get
\bea\label{large_dev_form}
P_{\rm an}(\tau) \approx \exp{\left[- \rho \mu {\cal T}_{\rm diff} H\left( \frac{\tau - {\cal T}_{\rm ad}}{{\cal T}_{\rm diff}{}}\right)\right]} \;.
\eea
We have to keep in mind that ${\cal T}_{\rm diff}$ is actually large, namely of order ${O}(\sqrt{L})$, and therefore, this form (\ref{large_dev_form}) should be considered as a large deviation form. To obtain the distribution of the typical fluctuations of $\tau$ from \eqref{large_dev_form} we need to center and scale properly the distribution. This is a rather common procedure in the context of extreme value statistics \cite{MPS2020}. To this purpose, let us define the function $a(\sigma)$ such that
\be \label{def_a}
\frac{2\sigma}{\sqrt{\pi} a^2(\sigma)}\ed^{- \frac{a^2(\sigma)}{4}} = 1 \;.
\ee
It turns out that $a(\sigma)$ can be expressed in terms of the $W$-Lambert function 
\be \label{a_Lambert}
a(\sigma) = 2\sqrt{W\left( \frac{\sigma}{2\sqrt{\pi}}\right)}
\ee
where we recall that the $W$-Lambert function is the reciprocal function of $f(w)= w \,\ed^{w}$. This form \eqref{a_Lambert} is particularly useful to obtain a precise asymptotic behavior of $a(\sigma)$ for large $\sigma$ \cite{WLambert}, which is the limit of interest here. To leading order, we have
\be \label{large_z}
a(\sigma) = 2\sqrt{\ln \sigma} + o(1) \;, \;\; {\rm as} \;\; \sigma \to \infty \;.
\ee
Then, from the scaling form given in \eqref{large_dev_form}, together with the asymptotic behavior of $H(z)$ for large $z$, we have, setting the dimensionless parameter $\sigma = \rho \mu {\cal T}_{\rm diff}$
\bea \label{Gumbel}
\lim_{L \to \infty} P_{\rm an}\left(\tau = {\cal T}_{\rm ad} + {\cal T}_{\rm diff}\left(a(\sigma)  + \frac{z}{a(\sigma)}  \right)\right)  = \ed^{-\ed^{-z}} \:,
\eea
which is the well known Gumbel distribution \cite{Gum58}. This relation \eqref{Gumbel} can equivalently be written as
\bea\label{proba_Gumbel}
T_{\rm vac} \overset{d}{=} {\cal T}_{\rm ad} + a(\sigma) {\cal T}_{\rm diff} + \frac{{\cal T}_{\rm diff}}{a(\sigma)} \, \gamma  \;, \; {\rm with} \;\; \sigma = \rho \mu {\cal T}_{\rm diff} = \rho \sqrt{\frac{DL}{\mu}} \;,\;\; {\rm as} \;\; L \to \infty \;,
\eea 
where $a(\sigma)$ is given in \eqref{a_Lambert} and $\gamma$ is a random variable which is distributed according to a Gumbel law, i.e., ${\rm Prob}(\gamma \leq z) = \ed^{-\ed^{-z}}$.

 \section{Conclusion and perspectives} 
 In this work we have studied different aspects of the last-passage time of linear diffusions. By using the mapping on a Schr\"{o}dinger problem we have identified the role of the Weyl coefficient and demonstrated its usefulness on several examples. It provides a complete description of the spectral properties of Dirichlet-Schr\"{o}dinger operators on the half line which is precisely the information which is needed in the study of last-passage times. It would be of great interest to extend this approach in higher dimension by using generalizations of the Weyl theory \cite{Amrein}.
 
 As an application of these results we have studied the emptying time of a box containing $N$ independent Brownian particles subject to a constant drift. The analysis of this problem involves statistical properties of independently but not identically distributed random variables which is a subject of current interest (see for instance \cite{LGMS2018,DDMS2017}). It would be very interesting to extend these results to other types of stochastic processes, such as run-and-tumble particles (RTP). Questions related to first \cite{Malakar,Dhar} and last \cite{Sing} passage times of a single RTP started only recently to be investigated and it would be natural to extend the computations performed here for the emptying time of a box to $N$ independent RTPs. 
 
 Finally the extension in higher dimension raises interesting problems  similar to those encountered in the study of the narrow escape problem, a topic which continues to draw significant interest for its biophysical applications \cite{Holcman}.

\vspace*{1cm}

\noindent{\bf Acknowledgments:} We acknowledge Michel Bauer, Eugene Bogomolny and Yves Tourigny for several useful remarks.
\appendix
\section{Transition kernel}
\label{KernelExpression}

The transition kernel $P(y,t \vert x,0)$ is related to the Schr\"odinger kernel by the well known formula (see for instance \cite{vanKampen1992}) 
\be
\label{relToSchroedinger}
P(y,t \vert x,0)=\ed^{-\frac{1}{2D}[U(y)-U(x)]} \langle y \vert \ed^{-tH} \vert x\rangle
\ee
where 
\be
\label{Hdefinition}
H=-D\frac{\dd^2}{\dd x^2} +\frac{1}{4D} \left[ \frac{\dd U}{\dd x} \right]^2 -\frac{1}{2}\frac{\dd^2 U}{\dd x^2}.
\ee
By Laplace transform one gets
\be
\widehat{P}_\lambda(y \vert x)= \ed^{-\frac{1}{2D}[U(y)-U(x)]} \langle y \vert \frac{1}{H+\lambda} \vert x\rangle.
\ee
Now, it is a standard property of the Sturm-Liouville problem that, for the values of $\lambda$ which are not in the spectrum of $H$, the resolvent can be expressed in terms of two independent solutions of the Schr\"odinger problem
\be
(H +\lambda) \psi_{-\lambda,R}=0
\ee
\be
(H +\lambda) \psi_{-\lambda,L}=0
\ee
which  vanish  in $+\infty$ and $-\infty$. Therefore 
\be
\widehat{P}_\lambda(y \vert x)= \ed^{-\frac{1}{2D}[U(y)-U(x)]}\frac{\psi_{-\lambda,L}(y) \psi_{-\lambda,R}(x)}{D \, W}
\qquad \textrm{if $y<x$} \;,
\ee
and 
\be
\widehat{P}_\lambda(y \vert x)= \ed^{-\frac{1}{2D}[U(y)-U(x)]}\frac{\psi_{-\lambda,L}(x) \psi_{-\lambda,R}(y)}{D \,  W}
\qquad \textrm{if $y>x$} \;,
\ee
where $W=\psi_{-\lambda,R}(x)\psi'_{-\lambda,L}(x)-\psi'_{-\lambda,R}(x)\psi_{-\lambda,L}(x)$ (with $\psi'= \dd \psi /\dd x$))  is the Wronskian which is independent of $x$. 

Let us set
\be
\label{relphiRpsiR}
\phi_\lambdaR(x) = \ed^{\frac{1}{2D} U(x)} \,\, \psi_{-\lambda,R}(x)
\ee
\be
\label{relphiLpsiL}
\phi_\lambdaL(x) = \ed^{\frac{1}{2D} U(x)} \,\, \psi_{-\lambda,L}(x) \;.
\ee
One can easily check that  $\phi_\lambdaR(x)$ and $\phi_\lambdaL(x)$ are two fundamental solutions satisfying \eqref{Generator}, $\Gen_x \phi_\lambda = \lambda \phi_\lambda$. One obtains
\be
\label{PversusphiW1}
\widehat{P}_\lambda(y\vert x) = \ed^{-\frac{1}{D}U(y)}\frac{\phi_\lambdaL(y) \phi_\lambdaR(x)}{D W}
\qquad \textrm{if $y<x$}
\ee
\be
\label{PversusphiW2}
\widehat{P}_\lambda(y\vert x) = \ed^{-\frac{1}{D}U(y)}\frac{\phi_\lambdaL(x) \phi_\lambdaR(y)}{D W}
\qquad \textrm{if $y>x$} \;.
\ee
By noticing that $\ed^{U(y)/D}W= \phi_\lambdaR(y) \phi'_\lambdaL(y) - \phi'_\lambdaR(y) \phi_\lambdaL(y)$ the latter expressions for the transition kernel can be rewritten only in terms of the functions $\phi_\lambdaL$ and $\phi_\lambdaR$ ; the corresponding expression is given in  \eqref{LTTransitionKernel}.

The boundary conditions obeyed by $\phi_\lambdaR$ and $\phi_\lambdaL$ are derived from those satisfied by the transition kernel. In the case of a diffusion along the whole real axis the transition kernel $P(y,t\vert x,0)$ vanishes when $\vert y- x\vert \to +\infty$, as well as its Laplace transform, and according to \eqref{PversusphiW1}
\be
\lim_{y\to -\infty} \ed^{- \frac{1}{D}U(y)}\phi_\lambdaL(y)=0
\ee
  while \eqref{PversusphiW2} leads to
 \be
\lim_{y\to +\infty} \ed^{- \frac{1}{D}U(y)}\phi_\lambdaR(y)=0  \;.
 \ee
In the case of a diffusion along the positive real half-axis with reflection at $y=0$ the probability current $J_P(y,t)$ (for a given initial position $x$) vanishes at $y=0$. The  expression for the probability current  $J_P(y,t)$ is given by the probability conservation $ \partial  J_P/\partial y=-\partial P/\partial t$  and the forward Fokker-Planck equation \eqref{ForwardFK} obeyed by the probability $P(y,t \vert x,0)$ for a given initial position, $\partial P/\partial t =\Gen^\dag_yP$, with the result 
\be
\label{DefJP}
J_P(y,t)=\left[ D \frac{\partial}{\partial y} + \frac{\dd U}{\dd y}\right] P(y,t) \;.
\ee
The relation also holds after Laplace transformation and, according to the expression \eqref{PversusphiW1}  for the transition kernel when $x>y=0,$ the vanishing of the probability current at $y=0$ is equivalent to
\be
\label{ReflexionCond}
\phi'_\lambdaL(0)=0 \;.
\ee

\section{Boundary with reflection}
\label{FormulaIntermediate}

The  Laplace transform of  $\Prob_y(T_{x_\star} >t)$  for $y\neq x_\star$ can be related to the time-derivative of the probability $\Prob_y(T_{x_\star} \leq t)$ by using an integration by parts with the result
\be
\int_0^{+\infty} \ed^{-\lambda t} \frac{\partial  }{\partial  t}\Prob_y(T_{x_\star} \leq t)\, \dd t
= \lambda \int_0^{+\infty}  \ed^{-\lambda t}\Prob_y(T_{x_\star} \leq t) \, \dd t
=1-\lambda \int_0^{+\infty}   \ed^{-\lambda t}\Prob_y(T_{x_\star}>t) \, \dd t
\ee
On the other hand the time derivative of the probability $\Prob_y(T_{x_\star} \leq t)$
is  the  probability density function for the first passage time and its
 Laplace transform can be expressed as an average over the trajectories
\be
\int_0^{+\infty} \ed^{-\lambda t} \frac{\partial }{\partial t} \Prob_y(T_{x_\star} \leq t) \, \dd t=
\mathbb{E}_y\left[ \ed^{-\lambda T_{x_\star}}\right]
\ee
As a consequence of the latter two relations the  Laplace transform of  $\partial \Prob_y(T_{x_\star} >t) /\partial y$ reads
\be
\label{relLT}
\int_0^{+\infty}  \ed^{-\lambda t}\frac{\partial}{\partial y}\Prob_y\left(  T_{x_\star}>t  \right) \, \dd t= 
-\frac{1}{\lambda} \frac{\partial}{\partial y}
\mathbb{E}_y\left[ \ed^{-\lambda T_{x_\star}}\right]
\ee
 The Laplace transform of $\partial \Prob_y\left( T_{x_\star} > t-u \right)/\partial y\vert_{y=0}$ can be obtained from  the relation \eqref{relLT} and the expression \eqref{EspTx0infxstar}  : the Laplace transform  reads for $y<x_\star$
\be
\int_0^{+\infty}  \ed^{-\lambda t}\frac{\partial}{\partial y}\Prob_y \left(T_{x_\star}>t \right)  \, \dd t
=-\frac{1}{\lambda}\frac{\phi'_\lambdaL(y)}{\phi_\lambdaL(x_\star)} 
\ee
In the case where $y=0$ is a reflecting boundary $\phi'_\lambdaL(0)=0$ and then
$\partial \Prob_y\left( T_{x_\star} > t-u \right)/\partial y\vert_{y=0}$  vanishes at $y=0$.

\section{Duality and time reversal}
\label{Duality}
We consider two diffusions $x(\tau)$ and $\widetilde{x}(\tau)$ whose generators  are respectively
\be
\label{Generator2}
\Gen_x =D \frac{\partial^2}{\partial x^2} + F(x) \frac{\partial }{\partial x}
\ee
\be
\label{Generator3}
\widetilde{\Gen}_x =D \frac{\partial^2}{\partial x^2} - F(x) \frac{\partial }{\partial x}
\ee
 where $F(x)=-\dd U/\dd x$. Denoting by $\phi_\lambdaR$ and  $\widetilde{\phi}_\lambdaR$ the corresponding fundamental solutions,
 one can prove that
 \be
 \widetilde{\phi}_\lambdaR = \ed^{-\frac{U(x)}{D}}\frac{\dd}{\dd x}\phi_\lambdaR
\qquad
\frac{\dd}{\dd x}\widetilde{\phi}_\lambdaR =\frac{\lambda}{D}\ed^{-\frac{U(x)}{D}}\phi_\lambdaR.
\ee
The corresponding logarithmic derivatives $\mR(\lambda;0) = \left.\frac{ \phi'_\lambdaR(x)}{\phi_\lambdaR(x)}\right\vert_{x=0}$,  
$\mRt(\lambda;0) = \left.\frac{\tilde{ \phi}'_\lambdaR(x)}{\widetilde{\phi}_\lambdaR(x)}\right\vert_{x=0}$  therefore satisfy
\be
\mR(\lambda;0)=\frac{\lambda}{D\mRt(\lambda;0)}.
\ee
When the potential is symmetric, we can  for simplicity set $m=2\mR$ and $\widetilde{m}=2\widetilde{m}_R$. The Laplace transforms of the probability density of the last-passage time at the origin are thus given by
\be
\int_0^{+\infty} \ed^{-\lambda t} \, \mathbb{E}_{0}\left[\ed^{-\lambda' g_{0, t}} \right]\, \dd t=\frac{1}{\lambda}
\frac{m(\lambda) }{m(\lambda+\lambda')}=D\frac{m(\lambda)}{\lambda}\frac{\widetilde{m}(\lambda+\lambda')}{\lambda+\lambda'}
\ee
\be
\int_0^{+\infty} \ed^{-\lambda t} \, \mathbb{E}_{0}\left[\ed^{-\lambda' \widetilde{g}_{0, t}} \right]\, \dd t=\frac{1}{\lambda}
\frac{\widetilde{m}(\lambda) }{\widetilde{m}(\lambda+\lambda')}=D\frac{\widetilde{m}(\lambda)}{\lambda}\frac{m(\lambda+\lambda')}{\lambda+\lambda'}.
\ee
Exchanging $\lambda$ and  $\lambda+\lambda'$ in the latter equation gives
\be
\int_0^{+\infty} \ed^{-\lambda t} \, \mathbb{E}_{0}\left[\ed^{-\lambda' g_{0, t}} \right]\, \dd t=\int_0^{+\infty} \ed^{-\lambda' t} \, \mathbb{E}_{0}\left[\ed^{-\lambda \widetilde{g}_{0, t}} \right]\, \dd t.
\ee
Therefore $h(t-u)f(u)=\tilde{h}(u)\tilde{f}(t-u)$. More intrinsically this means that $g_{0,t}$ and $\widetilde{g}_{0,t}$ have the same law.

\section{Another derivation for the mean value of the last-passage time when $x_\star=x_0=0$ }
\label{secVoros}
In this appendix, we present an alternative derivation of \eqref{RateMeanValueBBis} and \eqref{gmean00}. It is based on a set of remarkable identities \cite{Voros1} satisfied by the spectral determinant of the Schr\"odinger Hamiltonian 
\be
H=-\frac{\dd^2}{\dd x^2}+V(x)
\ee
defined on $\mathbb R^{+}$ with Neumann (+)  or Dirichlet (-) boundary conditions at the origin.
There are essentially two different prescriptions to define the spectral determinant. It can be defined either through the generalized  zeta function
\be
Z(s,\lambda)=\sum_{n\geq 0} \frac{1}{(E_n+\lambda)^s}
\ee
as 
\be
D(\lambda) = \ed^{-\frac{d}{ds} Z(s,\lambda)}\vert_{s=0}
\ee
or as an ordinary Fredholm determinant
\be
\Delta(\lambda)= \prod_{n \geq 0} \left(1+\frac{\lambda}{E_n}\right)
\ee
with an appropriate Weierstrass prefactor if necessary (see \cite{Voros1}).
One can also define the corresponding Neumann and Dirichlet determinants
\be
\label{Videntity3}
\Delta^{\pm}(\lambda) =\prod_{n \geq 0} \left(1+\frac{\lambda}{E_n^{\pm}}\right) \;.
\ee
where $E_n^+$  and $E_n^-$ denote the $n$th even and odd eigenvalues respectively.
Let us for simplicity restrict ourselves to the case where $\Delta (\lambda)$ is an entire function of order $\mu<1$. In this case (\cite{Voros2}) the two determinants $\Delta(\lambda)$ and $D(\lambda)$ just differ by a multiplicative constant independent of $\lambda$.
Moreover the Weyl function $\psi_{-\lambdaR} $ can be normalized in such a way that the following identities hold
\be
\label{Videntity}
\begin{cases}
\psi'_{-\lambdaR}(0)&= -D^{+} (\lambda)\\
\psi_{-\lambdaR}(0) &= D^{-} (\lambda) \;.
\end{cases}
\ee
Since $D^+$ and $\Delta^{+}$ just differ by a multiplicative constant, it follows that
\be
\label{Videntity2}
m_{R}(\lambda,0)=\frac{\psi'_{-\lambdaR}(0)}{\psi_{-\lambdaR}(0)}= C\frac{\Delta^{+} (\lambda)}{\Delta^{-}(\lambda)} \;.
\ee
By inserting  \eqref{Videntity2}  and \eqref{Videntity3} into \eqref{LTmeanValueUeven}  we get
\be
\label{LTmeanValueUevenBis_App}
\int_0^{+\infty} \ed^{-\lambda t} \, \mathbb{E}_{0}\left[g_{0, t} \right]\, \dd t
= \sum_n 
\left[\frac{1}{\lambda(\lambda +E_n^{+})} - \frac{1}{\lambda(\lambda +E_n^{-})} \right]
\;,
\ee
where the summations run over the even and odd states respectively. The contribution from the  ground state must be subtracted in order to perform  partial fraction decompositions,
\be
\int_0^{+\infty} \ed^{-\lambda t} \, \mathbb{E}_{0}\left[g_{0, t} \right]\, \dd t
= \frac{1}{\lambda^2} + \sum_{n\neq 0} 
\left[\frac{1}{E_n^+} \left(\frac{1}{\lambda} - \frac{1}{\lambda +E_n^+}\right)
-  \frac{1}{E_n^-} \left(\frac{1}{\lambda} - \frac{1}{\lambda +E_n^-}\right)\right]
 \;.
\ee
By inverse Laplace transform we get the mean value of the last-passage time
\be
\label{gmean01}
 \mathbb{E}_{0}\left[g_{0, t} \right]
 =t + \sum_{n\neq 0}\left[\frac{1-\ed^{-E_n^+ t}}{E_n^+} -\frac{1-\ed^{-E_n^- t}}{E_n^-}\right]
\ee
and its rate of variation reads
\be
\label{RateMeanValueBis}
\frac{\dd }{\dd t} \mathbb{E}_{0}\left[g_{0, t} \right] = \Tr_{+} \,\left( \ed^{-t H} \right) -   \Tr_{-} \,\left(\ed^{-t H} \right) \;,
\ee
where $\Tr_{\pm}$ denotes the trace over the even (odd) eigenfunctions of $H$. This formula coincides with the one given in Eq. \eqref{RateMeanValueBBis}, obtained in the text by another method.

\section{Quenched distribution of the emptying time}\label{app_quenched}

In this appendix, we perform the computation of the quenched average $P_{\rm qu}(\tau)$ as defined in Eq. \eqref{def_quen}. We restrict ourselves
to the case of a uniform initial distribution of the positions, i.e. $P(x_0^{(i)};t=0) = 1/L$. In this case, $P_{\rm qu}(\tau)$ reads 
\bea\label{def_quenched}
P_{\rm qu}(\tau) = \exp\left[ \int_0^L \frac{\dd x^{(1)}_{0}}{L} \cdots  \int_0^L \frac{\dd x^{(N)}_{0}}{L} \ln \Prob(T_{\rm vac} \leq \tau )\right] \;.
\eea
Using the fact that $T_{\rm vac}$ is the maximum among the $N$ random random variables $g_{L,\infty}^{(i)}$ [see Eq. \eqref{def_Tvac_max}], i.e. $\Prob (T_{\rm vac} \leq \tau ) = \prod_{i=1}^N {{\rm Prob}_{x_i}(g^{(i)}_{L,\infty} \leq \tau)}$, the argument of the exponential in (\ref{def_quenched}) reads 
\bea 
\int_0^L \frac{\dd x^{(1)}_{0}}{L} \cdots  \int_0^L \frac{\dd x^{(N)}_{0}}{L} \ln \Prob(T_{\rm vac} \leq \tau ) &=& \rho \int_0^L  \ln {\rm Prob}_x (g_{L,\infty} \leq \tau) \dd x \\
&=& \rho \int_0^L  \ln \left[1 - q(\tau,x) \right] \dd x \;, \label{qu_1}
\eea
where $q(\tau,x) = {\rm Prob}_x(g_{L,\infty} \geq \tau)$ [see \eqref{defqbar}]. We now focus on the scaling limit where $N \to \infty$, $L \to \infty$ keeping $N/L = \rho$ fixed. In this limit, one can replace $q(\tau,x)$ by its scaling form obtained in Eq. (\ref{scaling_qtaux}). As for the annealed average, the integral over $x$ in (\ref{qu_1}) is then dominated, for large $L$, by $x = {\cal O}( {\cal T}_{\rm diff}) = {\cal O}(\sqrt{L})$ and we find that $P_{\rm qu}(\tau)$ in (\ref{def_quenched}) takes a similar scaling form as $P_{\rm an}(\tau)$ in \eqref{large_dev_form}, namely
\bea \label{scaling_qu}
P_{\rm qu}(\tau) \approx \exp{\left[- \rho \mu {\cal T}_{\rm diff} H_{\rm qu}\left( \frac{\tau - {\cal T}_{\rm ad}}{{\cal T}_{\rm diff}{}}\right)\right]} \;,
\eea
where the scaling function $H_{\rm qu}(z)$ is given by
\bea \label{Hqu}
H_{\rm qu}(z) = -\int_0^\infty \ln \left( 1 - \frac{1}{2} {\rm erfc}\left(\frac{1}{2}(z+v)\right)\right) \dd v  = -\int_0^\infty \ln \left( \frac{1}{2} {\rm erfc}\left(\frac{-1}{2}(z+v)\right)\right) \dd v\;,
\eea
where we have used ${\rm erfc}(x) + {\rm erfc}(-x) = 2$. Interestingly, the same function $H_{\rm qu}(z)$ (up to scale factors) appears in the context of extreme eigenvalues in the Ginibre ensemble of random matrix theory \cite{LGMS2018} (it is denoted $\phi_I(z)$ in that paper), as well as in the context of current distribution of independent Brownian particles \cite{DG09}. In particular, its asymptotic  behaviors are given by (see Eq. (22) of Ref. \cite{LGMS2018})
\be\label{asympt_Hqu}
H_{\rm qu}(z) \sim \begin{cases}
&\dfrac{2}{\sqrt{\pi}z^2} \, \ed^{- \frac{z^2}{4}} \;, \; z \to +\infty \\
& \\
& \dfrac{1}{12} |z|^3 \;, \; \hspace*{1.2cm} z \to -\infty \;.
\end{cases}
\ee
Note that while the behaviour for $z \to +\infty$ coincides with the one of $H_{\rm an}(z)$ [see Eq. (\ref{asympt_Han})], the behaviour for $z \to -\infty$ is actually different. Since the behavior of $H_{\rm qu}(z)$ for $z \to + \infty$ coincides with the one of $H(z)$, the same analysis carried out in the text in Section \ref{sec:Gumbel_an} yields the large $L$ typical behavior of $P_{\rm qu}(\tau)$ 
\bea \label{Gumbel_qu}
\lim_{L \to \infty} P_{\rm qu}\left(\tau = {\cal T}_{\rm ad} + {\cal T}_{\rm diff}\left(a(\sigma)  + \frac{z}{a(\sigma)}  \right)\right)  = \ed^{-\ed^{-z}} \:, \; \; \sigma = \rho \mu {\cal T}_{\rm diff} \;,
\eea
where the function $a(\sigma)$ is given in \eqref{a_Lambert}, which yields the same result as in Eq. (\ref{Gumbel}) for $P_{\rm an}(\tau)$. This shows, as announced in the text, that the typical behaviors of $P_{\rm an}(\tau)$ and $P_{\rm qu}(\tau)$ are actually exactly the same in the scaling limit considered here where $N \to \infty$, $L \to \infty$ with $\rho  = N/L$ fixed.

\end{document}